\documentclass{article}

\usepackage{arxiv}

\usepackage[utf8]{inputenc} 
\usepackage[T1]{fontenc}    
\usepackage{hyperref}       
\usepackage{url}            
\usepackage{booktabs}       
\usepackage{amsfonts}       
\usepackage{nicefrac}       
\usepackage{microtype}      
\usepackage{lipsum}

\usepackage{graphicx}
\usepackage{xcolor}
\usepackage{amsmath}
\usepackage{tikz}
\usepackage{pgfplots}
\usepackage{subfig}
\usepackage{comment}
\usepackage{appendix}

\definecolor{bleudefrance}{rgb}{0.19, 0.55, 0.91}
\definecolor{bronze}{rgb}{0.8, 0.5, 0.2}

\def\malu#1{
{\color{black}#1}
}

\def\alex#1{
{\color{black}#1}
}

\def\ale#1{
{\color{black}#1}
}

\def\alv#1{
{\color{black}#1}
}

\title{Assessing the spatio-temporal spread of COVID-19 via compartmental models with diffusion in Italy, USA, and Brazil}

\author{
  Malú Grave \\
  Dept. of Civil Engineering\\
  COPPE/Federal University of Rio de Janeiro \\
  P.O. Box 68506, RJ 21945-970, Rio de Janeiro, Brazil \\
  \texttt{malugrave@nacad.ufrj.br} \\
   \And
   Alex Viguerie \\
  Department of Mathematics\\
  Gran Sasso Science Institute\\
  Viale Francesco Crispi 7, L'Aquila, AQ 67100, Italy \\
  \texttt{alexander.viguerie@gssi.it} \\
  \And
    Gabriel F. Barros \\
  Dept. of Civil Engineering\\
  COPPE/Federal University of Rio de Janeiro \\
  P.O. Box 68506, RJ 21945-970, Rio de Janeiro, Brazil \\
  \texttt{gabriel.barros@coc.ufrj.br} \\
     \And
 Alessandro Reali \\
  Dipartimento di Ingegneria Civile ed Architettura\\
  Università di Pavia \\
  Via Ferrata 3, Pavia, PV 27100, Italy \\
  \texttt{alereali@unipv.it} \\
   \And
   Alvaro L. G. A. Coutinho \\
  Dept. of Civil Engineering\\
  COPPE/Federal University of Rio de Janeiro \\
  P.O. Box 68506, RJ 21945-970, Rio de Janeiro, Brazil \\
  \texttt{alvaro@nacad.ufrj.br} \\
}

\begin{document}
\maketitle

\begin{abstract}

\alex{The outbreak of COVID-19 in 2020 has led to a surge in interest in the mathematical modeling of infectious diseases. Such models are usually defined as compartmental models, in which the population under study is divided into compartments based on qualitative characteristics, with different assumptions about the nature and rate of transfer across compartments. Though most commonly formulated as ordinary differential equation (ODE) models, in which the compartments depend only on time, recent works have also focused on partial differential equation (PDE) models, \ale{incorporating the variation of an epidemic in space. Such research on PDE models within a Susceptible, Infected, Exposed, Recovered, and Deceased (SEIRD) framework has led to promising results in reproducing COVID-19 contagion dynamics. In this paper, we assess the robustness of this modeling framework by considering different geometries over more extended periods than in other similar studies}. We first validate our code by reproducing previously shown results for Lombardy, Italy. \alv{We then focus on the U.S. state of Georgia and on the Brazilian state of Rio de Janeiro, one of the most impacted areas in the world.} Our results show good agreement with real-world epidemiological data in both time and space for all regions across major areas and across three different continents, suggesting that the modeling approach is both valid and robust.}
\end{abstract}

\section{Introduction}

The COVID-19 pandemic has caused an extraordinary global disruption, both in terms of human lives and economic damage. To limit the spread of contagion, governments have taken unprecedented measures. Such actions have included, but not been limited to, quarantines, curfews, lockdowns, and domestic and international travel suspensions. While undoubtedly effective and considered necessary by many experts, these measures also carry a high cost in terms of economic impact and mental and physical health, among others. In part, such restrictions are motivated by the lack of reliable data \ale{and models} on this disease's transmission in time and space, forcing cautious \ale{(and not always entirely rational)} responses from the authorities and population. More than ever, these events demonstrate the need for tools to predict the spatio-temporal dynamics of contagion. \ale{It is important to highlight that such tools should be not only accurate, but, given the difference in the quality of available data in distinct regions of the world (or even within the same country), they should also be robust with respect to the data used for parameter identification. In this context, model validation - also against data possibly coming from different countries - should play an important role.} 

The mathematical modeling of epidemics is a well-established field, and has been applied with success for many important cases. Examples of past applications include Dengue \cite{erickson2010dengue}, Zika \cite{dantas2018calibration}, HIV \cite{mukandavire2010global}, SARS \cite{dye2003modeling}, Malaria \cite {midekisa2012remote}, Ebola \cite {lekone2006statistical}, and \ale{many} others. However, the urgency of the COVID-19 pandemic has motivated the need for more research in this area, with several models for this pandemic outbreak being presented in recent months \cite{viguerie2020diffusion, viguerie2020simulating, arino2020simple, giordano2020modelling, carcione2020simulation,volpatto2020spreading, oliveira2021mathematical, jha2020bayesian, korolev2020identification, computation9020018}. 

Such models typically follow a \textit{compartmental} framework, in which the population under study is divided into compartments \ale{with} assumptions about the nature and time rate of transfer from one compartment to another \cite{brauer2019mathematical, viguerie2020simulating}. The large majority of the \ale{proposed} compartmental models are composed of a system of ordinary differential equations (ODEs) in time. Though such ODE models are simple to formulate, analyze, and solve numerically, they do not naturally account for individuals' movements from one region to another. To incorporate spatial variations, one may define regional compartments corresponding to different geographic areas and introduce coupling terms to account for movements across regions. Such approaches have also been applied to COVID-19 in, for example, \cite{gatto2020spread, giordano2020modelling, arandiga2020spatial}. ODE models may be used in combination with other approaches as well, such as with a neural network \cite{la2020epidemiological} or an agent-based computational framework \cite{zohdi2020agent, silva2020covid, gharakhanlou2020spatio}.  While all these approaches are effective, they lack the ability to model spatial transmission at the continuous level, which may have some advantages.

To this end, one may instead resort to a compartment model based on partial differential equations (PDEs). Such models are decidedly less common than ODE ones, in particular, due to their increased difficulty and time involved in both implementation and numerical solution. However, such an additional cost is counterbalanced by the fact that PDE models naturally (and accurately) incorporate spatial information, allowing for a continuous description of spatio-temporal dynamics, with the potential to account for geographical features, population heterogeneity, and multi-scale dynamics with relative ease. While quite uncommon, PDE models have been already used with success in the past to model epidemics \cite{keller2013numerical, kim1996galerkin, cantrell2004}.

\par 
\malu{For the specific case of COVID-19, PDE models were applied in \cite{viguerie2020diffusion, viguerie2020simulating,grave2020adaptive,jha2020bayesian}, in which a compartmental SEIRD model \textit{(susceptible, exposed, infected, recovered, deceased)} was used.} A modified version of this model, with the addition of a quarantined compartment, was applied in \cite{computation9020018}. \ale{Such modeling framework has shown promising results, being able to recreate observed data and make predictions with reasonable accuracy, but these successful applications have been so far obtained on a limited scale, with results restricted to single, specific regions. Therefore, despite the clearly high potential shown by this modeling approach, important questions remain regarding its \textit{robustness}. In particular, it is still to be proven the model capability to guarantee reliable results for a wide range of areas, which may exhibit significantly different geographical, behavioral, and demographic characteristics.} Furthermore, even if the model may demonstrate the ability to represent different regions accurately, it is unclear to what extent parameters must be specifically tuned to obtain accurate results for a given case. \ale{In fact, ideally, a robust model should be applicable to a wide range of settings, requiring relatively little specific parameter tuning to produce acceptable results for a given case.}

\par \ale{Within this context, the current work aims at evaluating the robustness of the model used in \cite{viguerie2020diffusion, viguerie2020simulating, grave2020adaptive} by studying in detail its application to a series of different cases. In order to accomplish this task, the model is applied to the following three cases: The Italian region of Lombardy, the U.S. state of Georgia, and the Brazilian state of Rio de Janeiro. These are not only the homelands of the authors, but they do represent regions of three distinct continents with very different features in terms of culture, population, and mobility, as well as government response to the pandemic. To deeply test its robustness to different region-specific characteristics, the model is applied across each case with limited parameter tuning and case-specific adjustments, and it is shown to be capable of producing consistently reliable results, in particular for predictions relative to short to mid-time spans, which are  relevant for timely political decisions.}

The remainder of this work is organized as follows: \ale{Section \ref{GE} introduces the governing equations and the relevant parameters, as well as the adopted terminology and notation. This section is also used to briefly discuss considerations regarding the numerical solution of the associated discrete problem. In the subsequent sections, the model is then tested on the cases of Lombardy, Georgia, and Rio de Janeiro, and for each case extensive discussion and analysis of results are provided.  Finally, the paper is concluded by a summary of the main findings and suggestions for future research directions.}

\section{Governing equations}\label{GE}

We present the governing equations following the continuum mechanics framework first shown in \cite{viguerie2020diffusion}, as opposed to more traditional notations common in mathematical and biological references. We consider a system that may be decomposed into $N$ distinct populations: $u_1(\mathbf{x}, t)$, $u_2(\mathbf{x}, t)$, ..., $u_N (\mathbf{x}, t)$.  Let $\Omega \in R^2$ be a simply connected domain of interest and denote its boundary as $\partial \Omega = \Gamma_D \cap \Gamma_N$. Let $[0, T]$ denote a generic time interval. The transient nonlinear diffusion-reaction system of equations written in compact vector notation then reads as follows:

\begin{equation}\label{eq:system}
    \frac{\partial\mathbf{u}}{\partial t} + \left(\mathbf{A} + \mathbf{B(u)}\right)\mathbf{u} - \nabla \cdot (\boldsymbol{\nu} \nabla\mathbf{u}) - \mathbf{f}=0 \textrm{ in }\Omega \times [0,T] 
\end{equation}

\begin{equation}\label{eq:direchlet}
\mathbf{u}=\mathbf{u_D} \textrm{ in }\Gamma_D \times [0,T] 
\end{equation}

\begin{equation}\label{eq:neumann}
(\boldsymbol{\nu} \nabla\mathbf{u}) \cdot \mathbf{n}=\mathbf{h} \textrm{ in }\Gamma_N \times [0,T], 
\end{equation}
\noindent where $s(\mathbf{x}, t)$, $e(\mathbf{x}, t)$, $i(\mathbf{x}, t)$, $r(\mathbf{x}, t)$, and $d(\mathbf{x}, t)$ denote the densities of the \textit{susceptible}, \textit{exposed}, \textit{infected}, \textit{recovered}, and \textit{deceased} populations, respectively. We additionally denote the \textit{cumulative number of infected} as $c(\mathbf{x}, t)$  and the sum of the living population as $n(\mathbf{x}, t)$ ; i.e.,
$n(\mathbf{x}, t) = s(\mathbf{x}, t) + e(\mathbf{x}, t) + i(\mathbf{x}, t) + r(\mathbf{x}, t)$. We then use the compact vector notation to represent these quantities as: $\mathbf{u} = [s,e,i,r,d]^T$. 
\par The tensors $\mathbf{A}$, $\mathbf{B}$ and $\boldsymbol{\nu}$, and the vector $\mathbf{f}$ are defined according to the particular dynamics of the system in question. In most applications, \alv{ $\boldsymbol{\nu}=\boldsymbol{\nu}(\mathbf{x,t})$, that is, diffusion is time-dependent, heterogeneous, and anisotropic;} however, this is not strictly necessary. In addition to the boundary  conditions \eqref{eq:direchlet}, \eqref{eq:neumann}, we must also specify an initial condition $\mathbf{u}(\mathbf{x},0)=\mathbf{u}_0$. We define the \textit{total population} $U_i(t)$ of a given compartment $u_i(\mathbf{x}, t)$ as:

\begin{equation}
    U_i(t) = \int_\Omega u_i(\mathbf{x}, t) d\Omega
\end{equation}

\noindent for $i=1,2, \cdots, N$.

To adequately define a model of COVID-19 contagion, we make the following assumptions, as discussed in \cite{viguerie2020diffusion, grave2020adaptive}:

\begin{itemize}
    \item We consider only mortality due to COVID-19;
    \item We do not consider new births;
    \item We assume that a portion of exposed individuals will never develop symptoms (asymptomatic cases), and hence will move directly from the exposed compartment to the recovered compartment;
    \item We assume that both pre-symptomatic and asymptomatic (the exposed compartment), as well as symptomatic (the infected compartment) individuals may spread the disease;
    \item There is a latency period after exposure and before the development of symptoms;
    \item Movement is proportional to the population size; i.e., we expect greater diffusion to occur in heavily populated regions;
    \item There is no movement among the deceased population.

\end{itemize}

Under the above assumptions, the \textit{frequency-dependent} system of equations reads:

\begin{equation}\begin{split}
&\frac{\partial s}{\partial t} + \frac{\beta_i}{n} si + \frac{\beta_e}{n} se 
- \nabla \cdot (n\nu_s\nabla s) = 0
\end{split}
\label{covid_s}
\end{equation}
\begin{equation}
\begin{split}
&\frac{\partial e}{\partial t} - \frac{\beta_i}{n} si - \frac{\beta_e}{n}se + (\alpha  + \gamma_e) e 
- \nabla \cdot (n\nu_e \nabla e) = 0
\end{split}
\label{covid_e}
\end{equation}
\begin{equation}
\frac{\partial i}{\partial t} - \alpha e + (\gamma_i  + \delta) i - \nabla \cdot (n\nu_i \nabla i) = 0
\label{covid_i}
\end{equation}
\begin{equation}
\frac{\partial r}{\partial t}  -\gamma_e e- \gamma_i i  -\nabla \cdot (n \nu_r \nabla r) = 0
\label{covid_r}
\end{equation}
\begin{equation}
\frac{\partial d}{\partial t} - \delta i = 0,
\label{covid_d}
\end{equation}

\noindent where 
$\beta_i$ and $\beta_e$ denote the transmission rates between symptomatic and susceptible individuals and asymptomatic and susceptible individuals, respectively (units days$^{-1}$), 
$\alpha$ denotes the incubation period (units days$^{-1}$), $\gamma_e$ corresponds to the asymptomatic recovery rate (units days$^{-1}$), $\gamma_i$ the symptomatic recovery rate (units days$^{-1}$), $\delta$ represents the mortality rate (units days$^{-1}$), and $\nu_s$, $\nu_e$, $\nu_i$, $\nu_r$ are the diffusion parameters of the different population groups as denoted by the sub-scripted letters (units km$^2$  persons$^{-1}$  days$^{-1}$). \alv{Note that all these parameters can be considered time and space-dependent.}



We may express the model (\ref{covid_s_dens})-(\ref{covid_d_dens}) in the general form given by equation \eqref{eq:system} by defining the tensors $\mathbf{A}$, $\mathbf{B}$, $\boldsymbol{\nu}$, and the vector $\mathbf{f}$ in the following way:

\begin{equation}
   \mathbf{A} = \begin{bmatrix}
0 & 0 & 0 & 0 & 0\\
0 & \alpha+\gamma_e & 0 & 0 & 0\\
0 & -\alpha & \gamma_i+\delta & 0 & 0\\
0 & -\gamma_e & -\gamma_i & 0 & 0\\
0 & 0 & -\delta & 0 & 0\\
\end{bmatrix}
\end{equation}
\begin{equation}
   \mathbf{B} = \begin{bmatrix}
0 & \frac{\beta_e}{n}s & \frac{\beta_i}{n}s & 0  & 0\\
0 & -\frac{\beta_e}{n}s & -\frac{\beta_i}{n}s & 0 & 0\\
0 & 0 & 0 & 0 & 0\\
0 & 0 & 0 & 0 & 0\\
0 & 0 & 0 & 0 & 0\\
\end{bmatrix}
\end{equation}
\begin{equation}
   \boldsymbol{\nu} = \begin{bmatrix}
\boldsymbol{\nu_s} & 0 & 0 & 0  & 0\\
0 & \boldsymbol{\nu_e} & 0 & 0 & 0\\
0 & 0 & \boldsymbol{\nu_i} & 0 & 0\\
0 & 0 & 0 & \boldsymbol{\nu_r} & 0\\
0 & 0 & 0 & 0 & 0\\
\end{bmatrix}
\end{equation}
\begin{equation}
   \boldsymbol{\nu_k} = \begin{bmatrix}
 \nu^k_{xx} & \nu^k_{xy}\\
 \nu^k_{yx} & \nu^k_{yy} 
\end{bmatrix}
\textnormal{ with } k=s,e,i,r
\end{equation}
\begin{equation}
   \mathbf{f} = \begin{bmatrix}
0\\
0 \\
0 \\
0 \\
0 \\
\end{bmatrix}.
\end{equation}


Assuming additionally that the region of interest is isolated, we then prescribe the following homogeneous Neumann boundary conditions:
 
 \begin{equation}
     \nabla s \cdot \mathbf{n} = 0
 \end{equation}
 \begin{equation}
     \nabla e \cdot \mathbf{n} = 0
 \end{equation}
  \begin{equation}
     \nabla i \cdot \mathbf{n} = 0
 \end{equation}
  \begin{equation}
     \nabla r \cdot \mathbf{n} = 0,
 \end{equation}
 
\noindent or simply $(\boldsymbol{\nu} \cdot \nabla\mathbf{u}) \cdot \mathbf{n} = 0$.

We note that one may also consider the following model related to (\ref{covid_s})-(\ref{covid_d}):

\begin{equation}\begin{split}
&\frac{\partial s}{\partial t} + \beta_i \left(1-\frac{A}{n}\right)si + \beta_e \left(1-\frac{A}{n}\right)se - \nabla \cdot (n\nu_s\nabla s) = 0
\end{split}
\label{covid_s_dens}
\end{equation}
\begin{equation}
\begin{split}
&\frac{\partial e}{\partial t} - \beta_i \left(1-\frac{A}{n}\right)si - \beta_e \left(1-\frac{A}{n}\right)se + (\alpha  + \gamma_e) e - \nabla \cdot (n\nu_e \nabla e) = 0
\end{split}
\label{covid_e_dens}
\end{equation}
\begin{equation}
\frac{\partial i}{\partial t} - \alpha e + (\gamma_i  + \delta) i - \nabla \cdot (n\nu_i \nabla i) = 0
\label{covid_i_dens}
\end{equation}
\begin{equation}
\frac{\partial r}{\partial t}  -\gamma_e e- \gamma_i i  -\nabla \cdot (n \nu_r \nabla r) = 0
\label{covid_r_dens}
\end{equation}
\begin{equation}
\frac{\partial d}{\partial t} - \delta i = 0.
\label{covid_d_dens}
\end{equation}
The model (\ref{covid_s_dens})-(\ref{covid_d_dens}) is known as the \textit{density-dependent} formulation and was studied in \cite{viguerie2020simulating,viguerie2020diffusion}. The difference between the models can be seen in equations (\ref{covid_s})-(\ref{covid_d}) and (\ref{covid_s_dens})-(\ref{covid_d_dens}). In (\ref{covid_s})-(\ref{covid_d}), the contact terms are normalized by the living population, whereas such a normalization does not occur in (\ref{covid_s_dens})-(\ref{covid_d_dens}). In the \textit{frequency-dependent} formulation (\ref{covid_s})-(\ref{covid_d}), this normalization implies that the contagion is independent of population density, while this is not the case in the density-dependent formulation, as the name may suggest \cite{oli2006population, thrall1995frequency}. Both models may be valid and deliver accurate results, depending on the physical situation, and we will show computations performed with both formulations in the present work. 
\par \alex{The other major difference between (\ref{covid_s})-(\ref{covid_d}) and (\ref{covid_s_dens})-(\ref{covid_d_dens}) is the presence of the Allee effect $A$. This term has been used extensively in other settings, with the form used above inspired directly by applications in cancer modeling \cite{johnson2019cancer, delitala2020Cancer}. The Allee term serves to reduce transmission in areas where the population density is below a certain threshold $A$, by bringing the population in the exposed compartment to the susceptible compartment in such regions. Consequently, in those areas, the population in compartments $e$ and $i$ tends to zero, which eventually cancels out the transfer term where the Allee term participates. In \cite{viguerie2020simulating}, this term served to accentuate the contagion in major urban regions while reducing it in less-populated areas, consistent with the observed physics. We briefly note also that as $s$, $i$, and $e$ are all less than $n$ by definition, we do not expect blowup of this term, even for very small $n$.}

\alex{For the numerical solution of (\ref{covid_s})-(\ref{covid_d}), (\ref{covid_s_dens})-(\ref{covid_d_dens}), we discretize in space using a Galerkin finite element formulation. The resulting systems of equations are stiff, leading us to employ implicit methods for time integration. We apply the second-order Backward Differentiation Formula (BDF2) in all cases, which offers second-order accuracy while remaining unconditionally stable. We additionally make use of an adaptive mesh refinement and coarsening strategy (AMR/C), allowing us to resolve multiple scales. One may find more details about the adopted methods in \cite{viguerie2020diffusion, grave2020adaptive}.}

\malu{All simulations have been performed using the \texttt{libMesh}, a C++ FEM open-source software library for parallel adaptive finite element applications \cite{libmesh}. \texttt{libMesh} also interfaces with external solver packages like PETSc \cite{petsc-web-page} and Trilinos \cite{trilinos-website}. It is an excellent tool for programming the finite element method and can be used for one-, two-, and three-dimensional steady and transient simulations.  This library provides native support for AMR/C, thus providing a natural environment for the present study. The main advantage of \texttt{libMesh} is the possibility of focusing on implementing the specific features of the modeling without worrying about adaptivity and code parallelization. Consequently, the effort to build a high performance computing code tends to be minimized.}

\section{Simulation of the COVID-19 virus spread in some of the world's hot-spots}

\par \alex{ In the following section, we demonstrate the robustness of the models (\ref{covid_s})-(\ref{covid_d}), (\ref{covid_s_dens})-(\ref{covid_d_dens}) by simulating the COVID-19 outbreak for three different geographic regions: Lombardy (Italy), \alv{the U.S. state of Georgia, and the state of Rio de Janeiro (Brazil).} These regions cover three different continents, and each exhibits very different geographical features, population characteristics, economic, social, cultural factors, and government responses. By achieving good results for each case, we aim at showing that the modeling approach shown here is robust to general cases and does not require extensive modification or specific parameter tuning to achieve reasonable results for a given case.}
\par \alex{The three cases shown were chosen for a number of reasons. Notably, they represent the authors' home regions, so there is an obvious interest from this point-of-view. Beyond this point, however, the regions also have specific characteristics that make them remarkable case studies. Italy is one of the hardest-hit regions in Europe, and Lombardy represented the epicenter of the outbreak in Italy, as well as the first area in a Western country to exhibit community spread. During the first wave of the epidemic in Spring 2020, Lombardy was disproportionately impacted compared to other regions in Italy, in which the epidemic was less severe. It remains the hardest-hit region in Italy, even after the second wave of infections has been less geographically restricted. As Lombardy was studied specifically in \cite{viguerie2020simulating, viguerie2020diffusion}, it also represents a natural choice for validating the new code and the adaptive-meshing strategy implemented in the present work.}
\par \alex{The U.S. state of Georgia is a logical choice for several reasons. The state has 159 counties, second to only Texas among U.S. states. However, the average county size for Georgia is less than half of Texas, giving the state a good county-by-county spatial resolution, and allowing us to evaluate the spatial accuracy of the model in reasonable detail. Further, the state was the first among U.S. states to reopen in late April, and has since taken a large number of policy measures, in both relaxing and increasing restrictions, providing us a large number of decisions to incorporate and analyze in our model. }
\par \alex{Rio de Janeiro is also a sensible choice for many of the same reasons as Georgia.} \malu{At the time of writing, Brazil is the third-most severely impacted country in the world in terms of total cases, and the second one in terms of total deaths. Rio de Janeiro has the largest number of deaths per inhabitant within Brazil, making its simulation worthwhile on severity alone.} \alex{ Aside from this aspect, the area also exhibits interesting geographical features, notably its natural harbor and large population centers on either side. Additionally, there is good availability of mobility data, which we have incorporated into the diffusion parameters. The use of such data is logical for a model of this type, and its consideration represents an interesting and important component for this modeling approach.}

\subsection{Lombardy, Italy}\label{AMR}

\par \alex{In our first numerical test, we seek to reproduce, and possibly improve, the simulations of Lombardy, Italy, as shown in \cite{viguerie2020simulating, viguerie2020diffusion}. This model was already shown in \cite{viguerie2020simulating} to predict the first wave of the COVID-19 outbreak in Lombardy with good accuracy. We note that for this simulation we use the density-dependent model with the Allee term (\ref{covid_s_dens})-(\ref{covid_d_dens}). While using the same parameters as shown in \cite{viguerie2020simulating, viguerie2020diffusion}, we incorporate two major improvements. First, we employ the adaptive mesh refinement and coarsening strategy shown in \cite{grave2020adaptive}. Second, rather than using Backward-Euler time-stepping, we utilize a BDF2 scheme, as \cite{Grave_Camata_Coutinho_2020, viguerie2020diffusion} suggest that such a scheme is better suited to the system of equations in question. Thus, the primary motivation of this simulation is the validation of the current code and algorithmic framework.} 

\malu{The AMR/C procedure uses a local error estimator to drive the refinement and coarsening procedure, considering the error of an element relative to its neighbor elements in the mesh. The elements are flagged  depending on a refining ($r_f$) and a coarsening ($c_f$) fraction. The refinement level is limited by a maximum $h$-level
($h_{max}$). In this case, for the AMR/C procedure, we set $h_{max}=1$, $r_f=0.95$, $c_f=0.01$. We apply the adaptive mesh refinement every 4 time-steps. The original mesh has 10,987 linear triangular elements, and, after refinement, the minimum spatial resolution is about 1 kilometer. We initially refine the domain in one level, and the time-step is defined as $\Delta t = 0.25$ $days$.}

\begin{figure}[htpb]
    \centering
    \includegraphics[width=\linewidth]{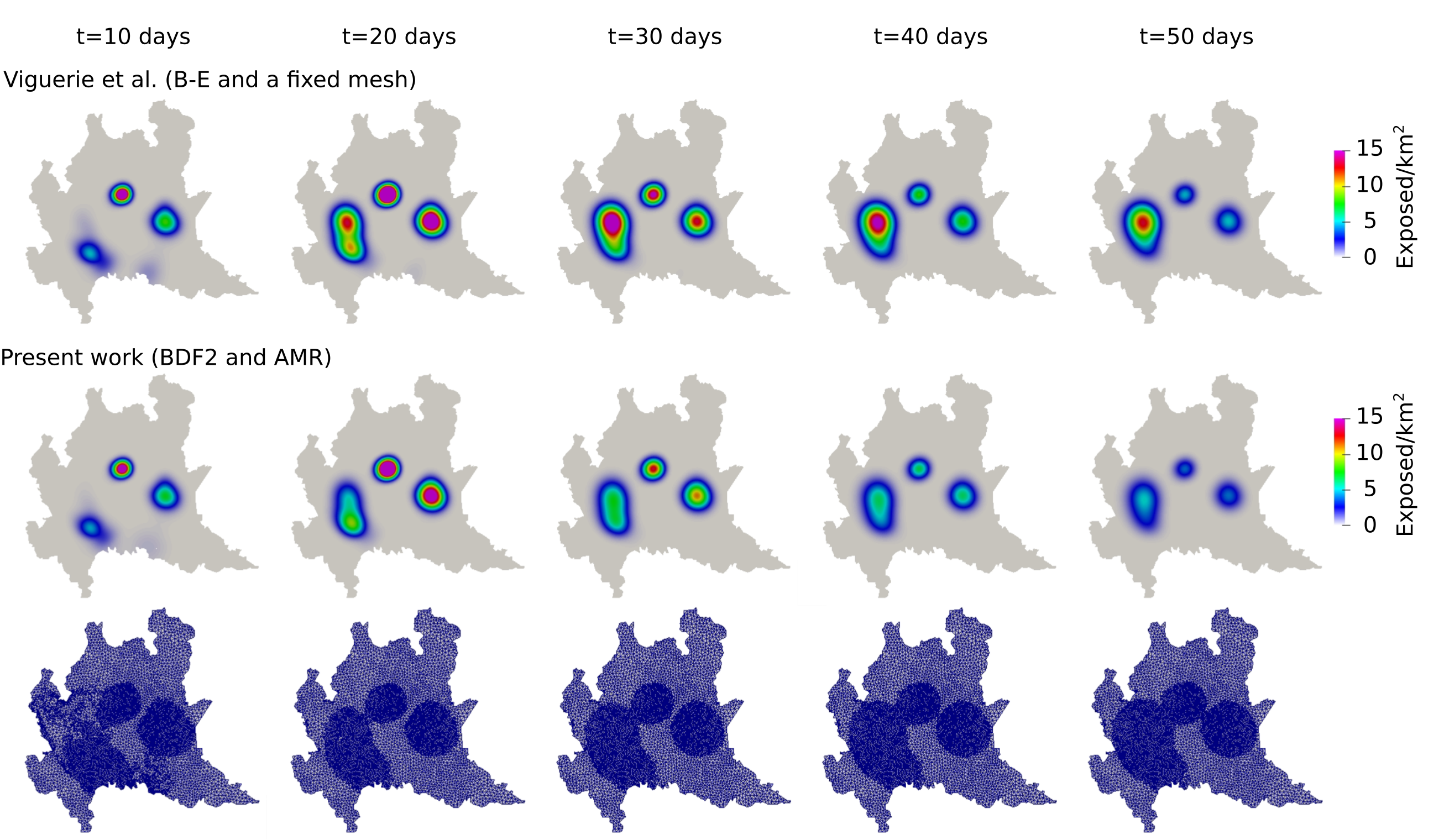}
    \caption{Spread of the COVID-19 virus spread at Lombardy, Italy. Comparison between the results obtained in \cite{viguerie2020simulating}, in which the time integration was done using the Backward-Euler method with a fixed mesh, and the present work, in which we use the BDF2 method and AMR. }
    \label{fig:Lombardy_comparison}
\end{figure}
\par \alex{In Fig. \ref{fig:Lombardy_comparison} we compare the results obtained here with those from \cite{viguerie2020diffusion}. We observe the same qualitative dynamics; the epidemic follows the same path, moving from the southern areas of the region northward into the major metropolitan areas, and the Milan area in particular. The changes to the meshing and time-stepping schemes result in somewhat lower contagion, particularly in the areas in and around the city of Milan in the western part of the region. Examining the progression of the mesh adaptation in Fig. \ref{fig:Lombardy_comparison}, we see that the AMR/C algorithm successfully follows the transmission path, changing in time to adapt to the evolving geography of the epidemic in time. We briefly note that when using the current code with Backward-Euler and a fixed mesh (not shown), the results were identical to those obtained in \cite{viguerie2020simulating}; thus, we can conclude that the differences between the solution depicted in the present work and that shown in \cite{viguerie2020simulating} represent improvements resulting from the differences in meshing and time-stepping used in the present one. From this, we conclude that the current code and algorithmic framework is validated.} 


\subsection{Georgia, USA}\label{Georgia_section}

\malu{From now on we use the frequency-dependent model (\ref{covid_s})-(\ref{covid_d}) together with adaptive mesh refinement and the BDF2 time-stepping scheme. In the next test, we aim at reproducing the COVID-19 outbreak}\alex{ in the U.S. state of Georgia}.

\subsubsection{Model Construction}

We have first obtained the map of the state of Georgia (GA) along with the county boundaries in shapefile format from 
\url{https://arc-garc.opendata.arcgis.com/datasets/dc20713282734a73abe990995de40497_68}. To triangulate the GA region, we follow the steps given by \cite{jha2020bayesian}:

\begin{itemize}
    \item Load the GA map file in \alex{ the freely available QGIS software \cite{QGIS_software}};
    \item Coarse grain the outer boundary segments using \alex{the} Simplify tool in QGIS. The original map has \alex{some} very small length segments, which may create problems in triangulation or result in a very fine mesh;
    \item Obtain the vertices using \alex{the} Extract Vertices tool in QGIS and save the vertices layer using \alex{the save layer as
option.} Select As XY in \alex{the} Graphical category \alex{and save} the file in a .csv format;
\item Prepare a Gmsh input file using the vertices' file for triangulation.
\end{itemize}

\begin{figure}[htpb]
    \centering
    \includegraphics[width=0.7\linewidth]{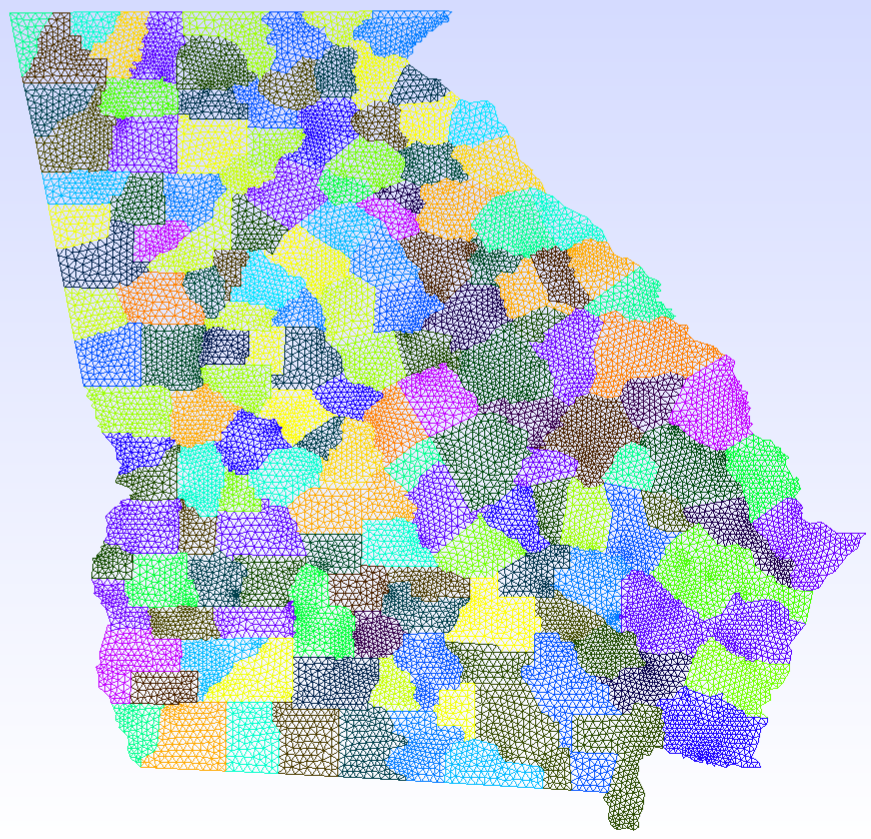}
    \caption{Map of the state of GA partitioned into 159 internal counties.}
    \label{fig:GAmesh}
\end{figure}

In Figure \ref{fig:GAmesh}, we show the generated grid \alex{for} the state of Georgia. The original mesh has 32,056 linear triangular elements, and after refinement, the minimum spatial resolution is about 2 kilometers. We initially refine the domain in one level. For the AMR/C procedure, we set $h_{max}=1$, $r_f=0.95$, $c_f=0.01$. We apply the adaptive mesh refinement every 4 time-steps. The time-step is defined as $\Delta t = 0.25$ $days$.


\begin{figure}[htpb]
    \centering
    \includegraphics[width=0.7\linewidth]{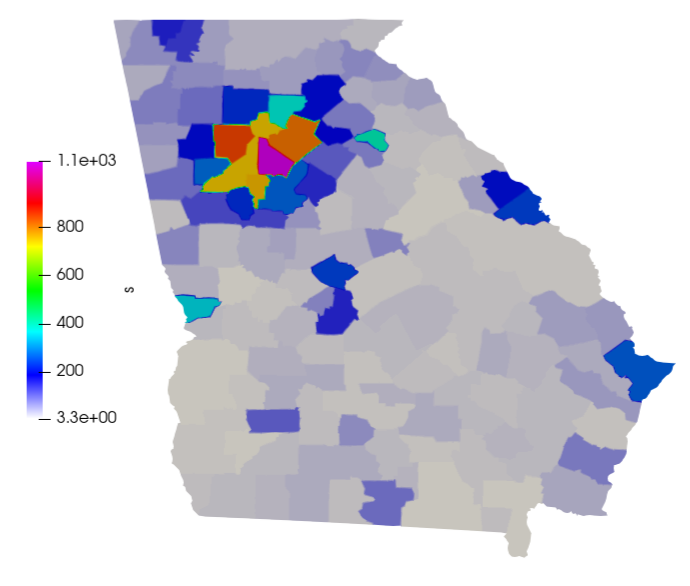}
    \caption{Initial susceptible population ($people/km^2$).}
    \label{fig:GAs}
\end{figure}

\begin{figure}[htpb]
    \centering
    \includegraphics[width=0.7\linewidth]{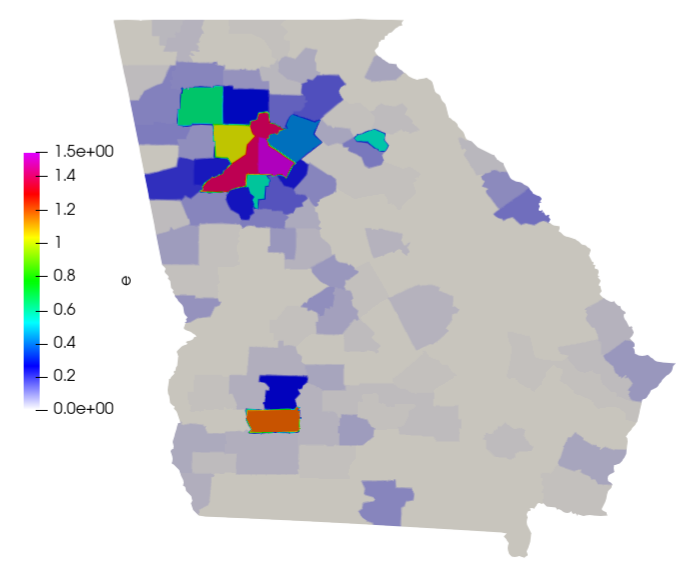}
    \caption{Initial exposed population ($people/km^2$).}
    \label{fig:GAe}
\end{figure}

\begin{figure}[htpb]
    \centering
    \includegraphics[width=0.7\linewidth]{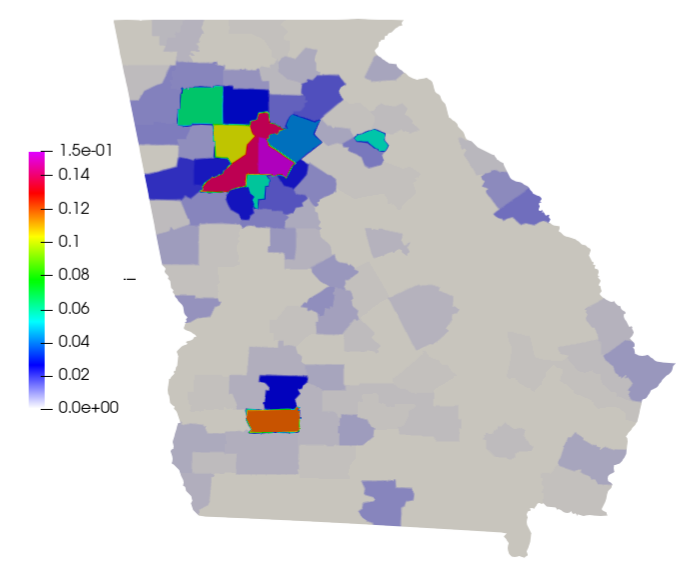}
    \caption{Initial infected population ($people/km^2$).}
    \label{fig:GAi}
\end{figure}

We define the initial infected population accordingly to the data provided by the \alex{Johns Hopkins University \cite{dong2020interactive}}. We define the beginning of the simulation on 25 March 2020 and simulate 250 days. The exposed population is more difficult to estimate, as they consist of asymptomatic and pre-symptomatic individuals. Here, we consider 10 times the number of infected \cite{reis2020characterization}. The susceptible population is based on the estimation of the population of each county, given by \alex{the U.S. Census Bureau\cite{GACensus}.} The populations are divided by the area of each county and distributed on the 159 areas of the mesh as $people/km^2$.
In Figs. \ref{fig:GAs}, \ref{fig:GAe}, and \ref{fig:GAi} we show the initial conditions.

The parameters of the simulation are defined as: $\alpha = 1/8$ $ day^{-1}$, $\gamma_i = 1/24$  $ day^{-1}$,  $\gamma_e = 1/6$  $ day^{-1}$, $\delta = 1/160$  $ day^{-1}$. These values are based on available data from the literature regarding the mortality, incubation period, and recovery time for infected and asymptomatic patients \cite{viguerie2020simulating}. 

\begin{figure}[htpb]
    \centering
    \includegraphics[width=\linewidth]{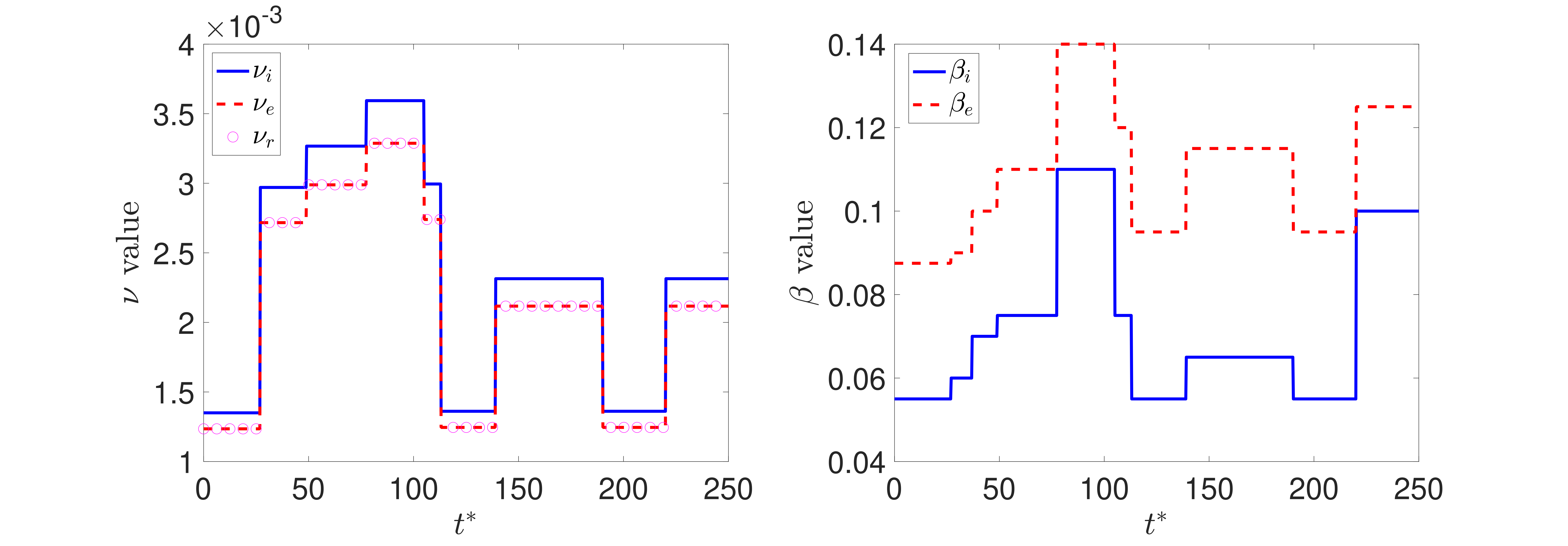}
    \caption{Diffusion (left) and contact rate (right) in time for Georgia.}
    \label{fig:GA_beta}
\end{figure}

The contact rate and diffusion are estimated based on the policies adopted in the Georgia state during the period of the simulation (\url{https://coronavirus.jhu.edu/data/state-timeline/new-deaths/georgia/}). This leads to the \alv{time-dependent} values as shown in Fig. \ref{fig:GA_beta}.

\subsubsection{Results}

\begin{figure}[htpb]
  \centering
    \subfloat{\includegraphics[width=0.45\textwidth]{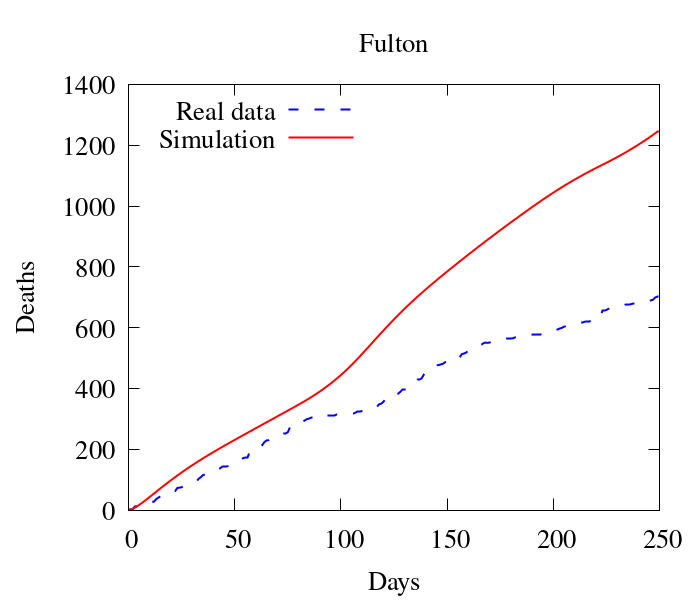}\label{Fulton}}
      \hfill
    \subfloat{\includegraphics[width=0.45\textwidth]{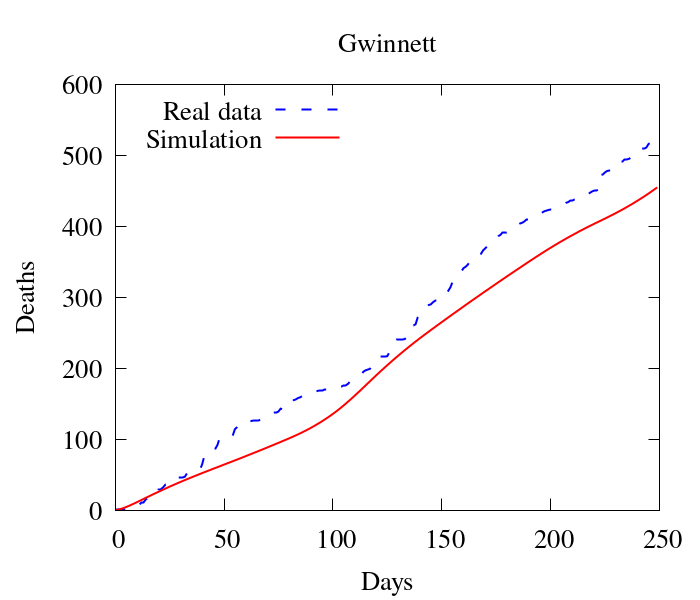}\label{Gwinnet}}
     \hfill
    \subfloat{\includegraphics[width=0.45\textwidth]{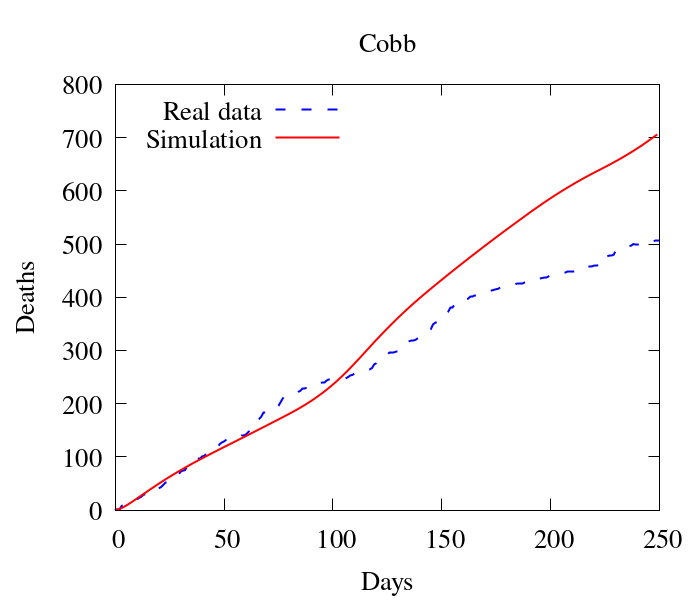}\label{Cobb}}
    \hfill
    \subfloat{\includegraphics[width=0.45\textwidth]{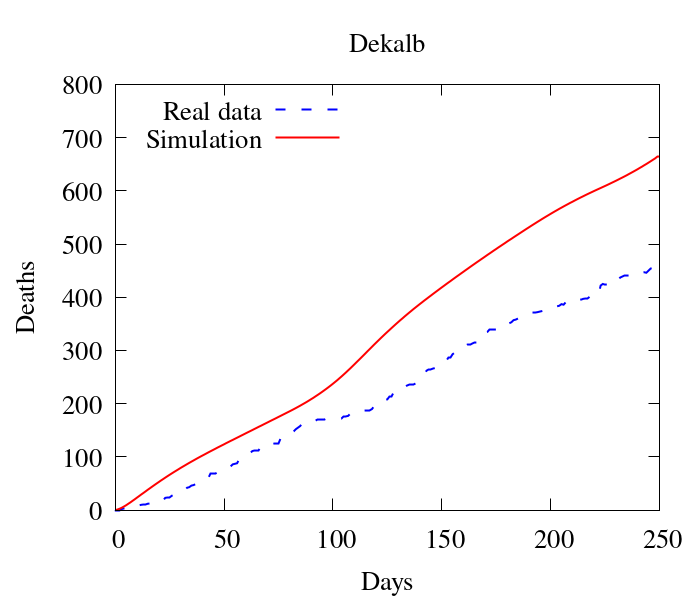}\label{Dekalb}}
    \hfill
    \subfloat{\includegraphics[width=0.45\textwidth]{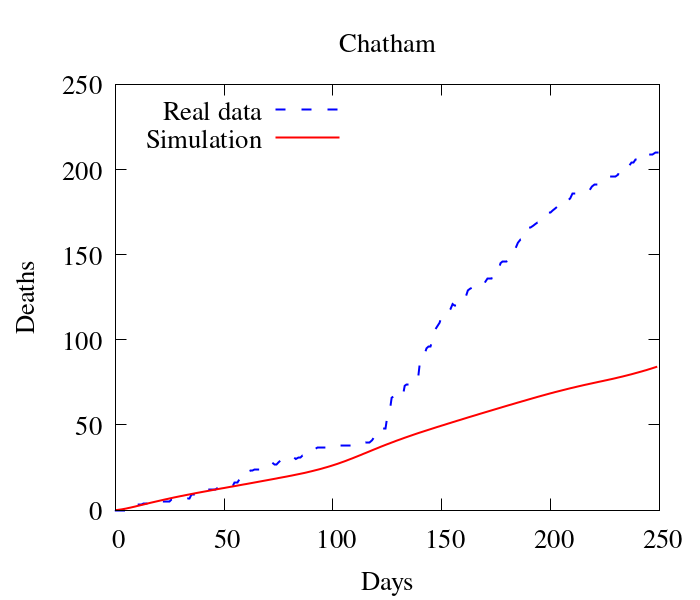}\label{Chatham}}
    \hfill
    \subfloat{\includegraphics[width=0.45\textwidth]{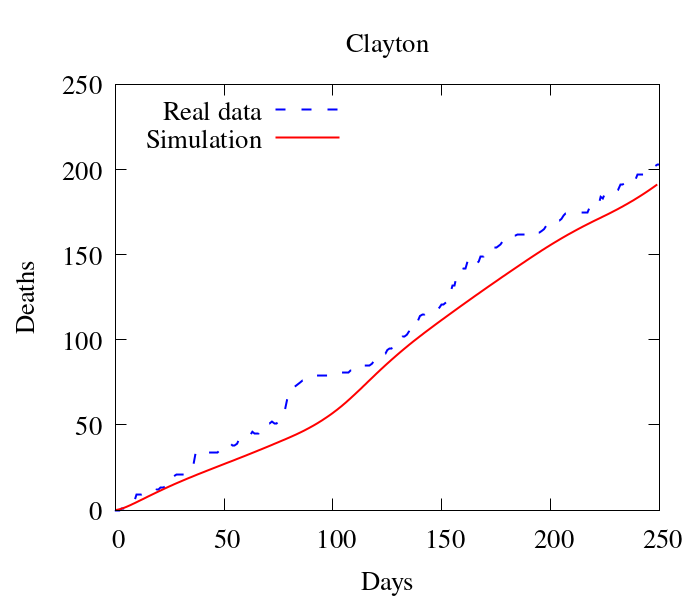}\label{Clayton}}
      \hfill
  \caption{Comparison between simulation and real data of deaths at Georgia (per county) - part 1.}
  \label{comp_Georgia_1}
\end{figure}

\begin{figure}[htpb]
  \centering
  \subfloat{\includegraphics[width=0.45\textwidth]{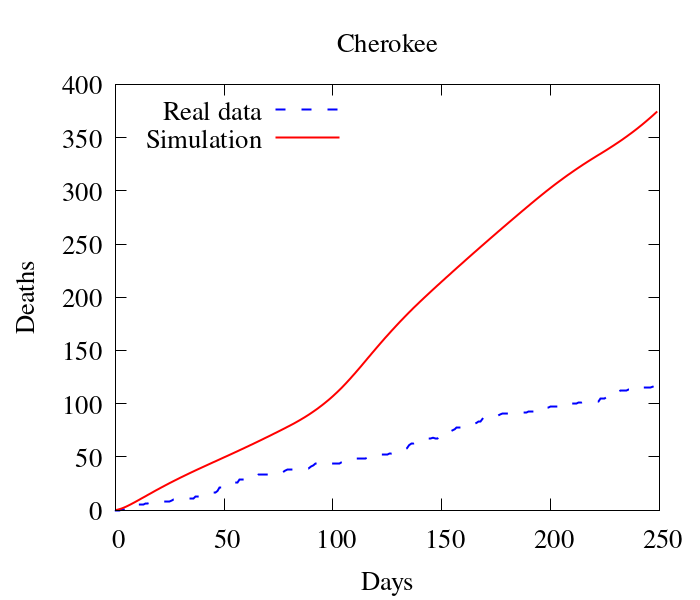}\label{Cherokee}}
    \hfill  
  \subfloat{\includegraphics[width=0.45\textwidth]{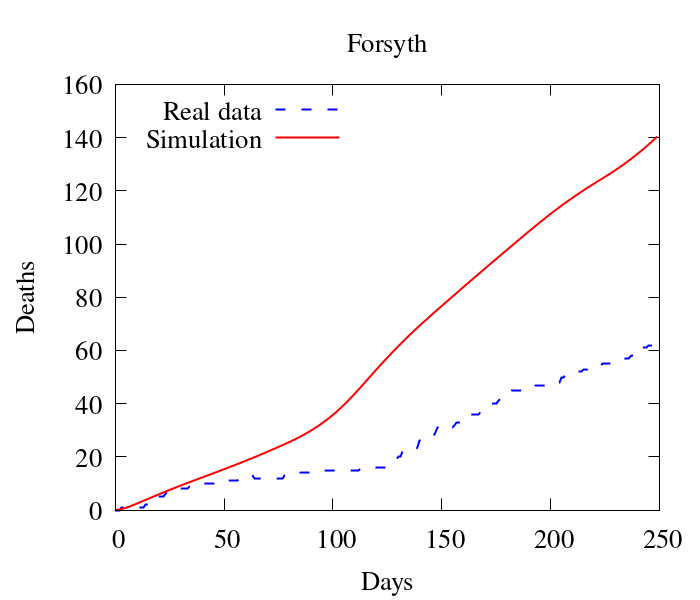}\label{Forsyth}}
    \hfill
  \subfloat{\includegraphics[width=0.45\textwidth]{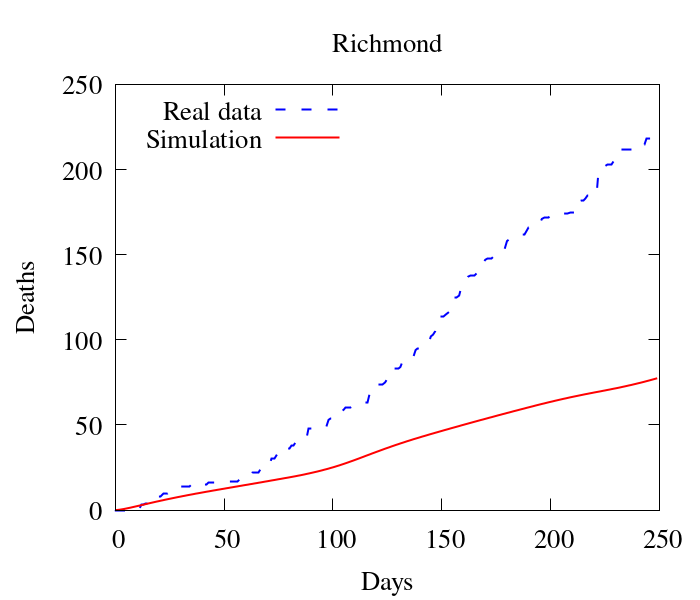}\label{Richmond}}
  \subfloat{\includegraphics[width=0.45\textwidth]{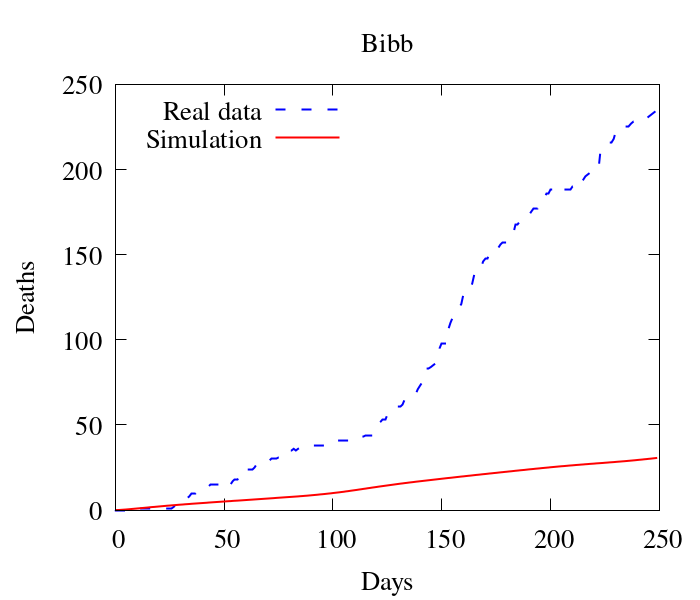}\label{Bibb}}
  \hfill
  \subfloat{\includegraphics[width=0.45\textwidth]{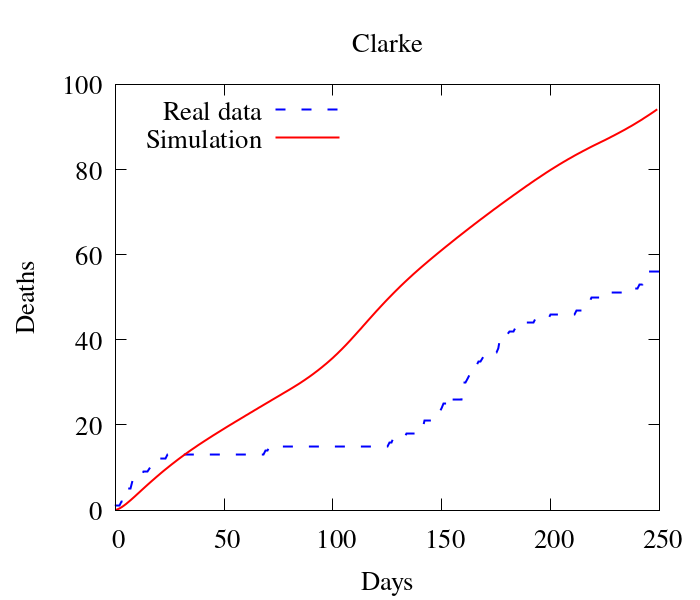}\label{Clarke}}
  \subfloat{\includegraphics[width=0.45\textwidth]{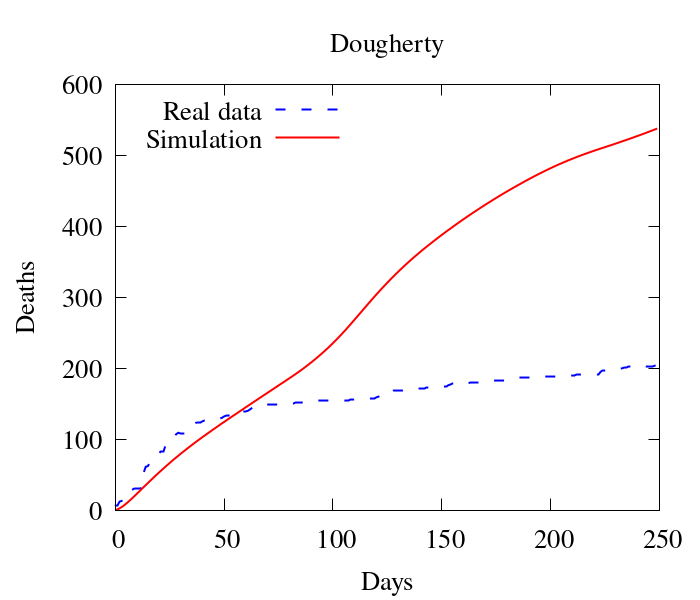}\label{Dougherty}}
    \hfill
  \caption{Comparison between simulation and real data of deaths at Georgia (per county) - part 2.}
  \label{comp_Georgia_2}
\end{figure}

\begin{figure}[htpb]
    \centering
    \includegraphics[width=\linewidth]{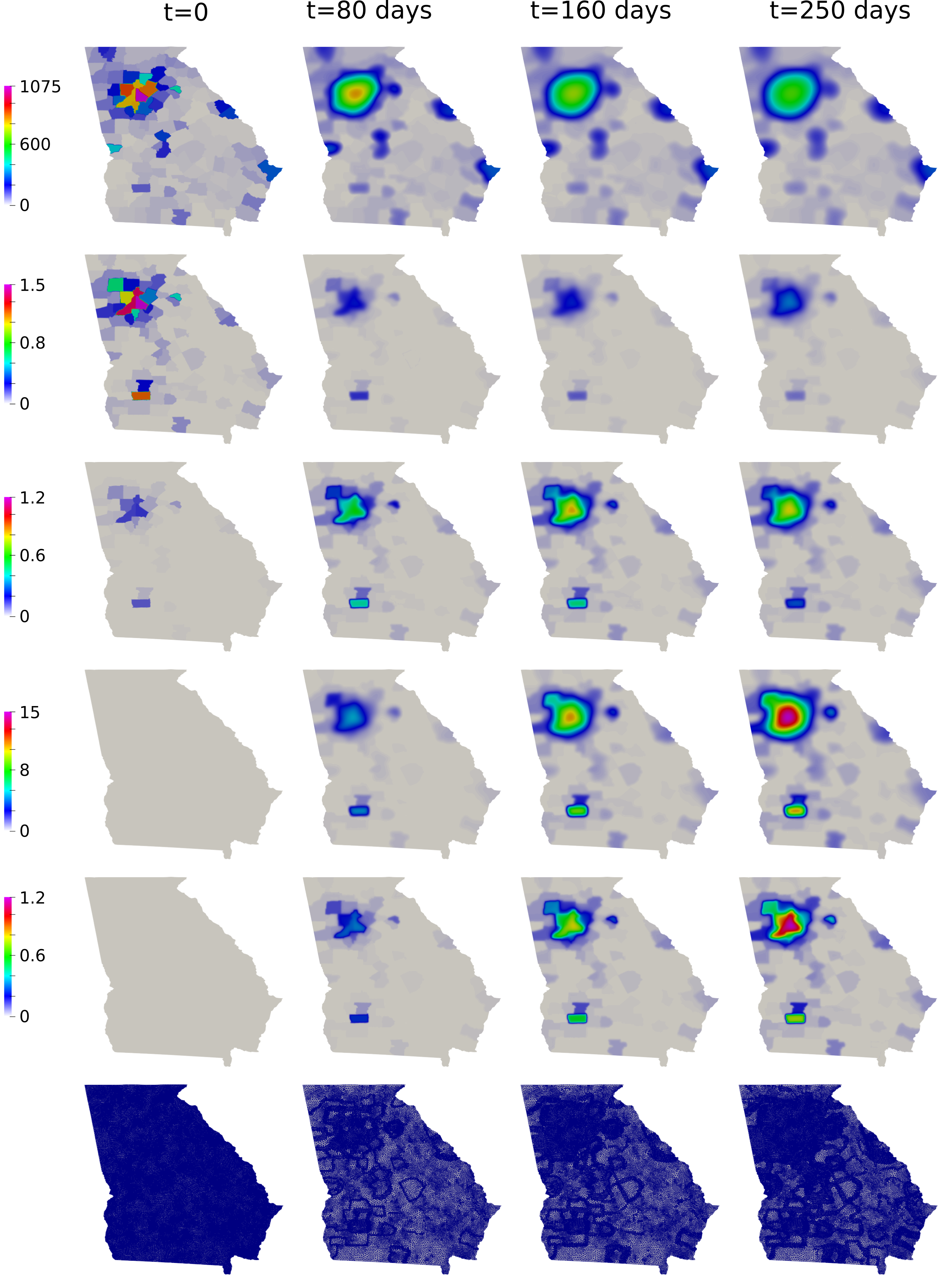}
    \caption{Snapshots of Georgia simulation. Top to bottom: Susceptible, Exposed, Infected, Recovery and Deceased population ($people/km^2$), followed by the mesh at different time-steps.}
    \label{fig:georgia_simulation}
\end{figure}

To compare our model results with the data available, we compare the number of deaths caused by the COVID-19. First, we show the values of the whole state. Then, we integrate the values of some counties and plot them in time (Fig. \ref{comp_Georgia_1} and \ref{comp_Georgia_2}). Fig. \ref{fig:georgia_simulation} also shows some snapshots of the simulation with the populations and mesh at different time-steps.

\par
\begin{figure}[htpb]
    \centering
    \includegraphics[width=0.6\linewidth]{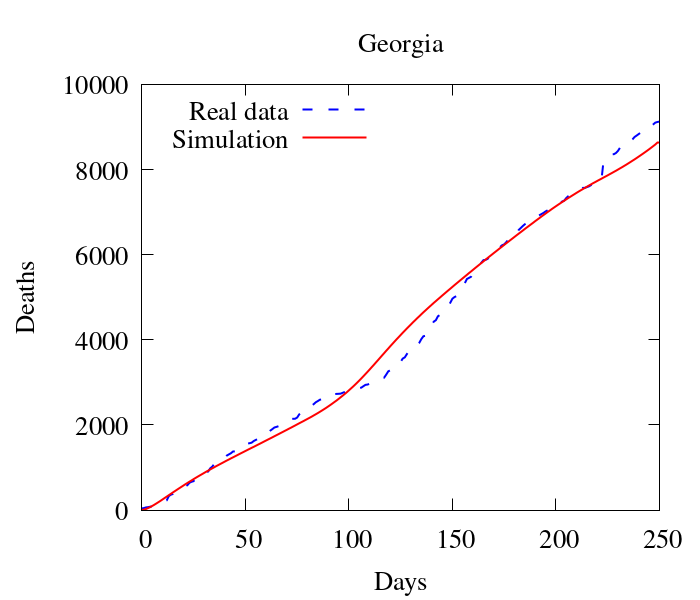}
    \caption{Comparison between simulation and real data of deaths at Georgia (total).}
    \label{Georgia}
\end{figure}

 Referring to the entire state in Fig. \ref{Georgia}, we observe that the model accurately captures both qualitative and quantitative dynamics of the epidemic with good accuracy. The difference in relative $L^2$ norm between the measured and simulated deaths is 5.75\%, confirming the agreement. We note that a small discrepancy between the simulated and measured data occurs at the end of the simulated period; this is caused by a series of deaths being added retroactively to the data, with the actual date of these deaths being unknown. 
\par Turning our attention to the individual counties, we see more variation in the results. We start by noting that the initial dynamics (first 60 days) is predicted well nearly everywhere, with a divergence between the simulated and measured deaths becoming more apparent as the epidemic progresses. This divergence generally indicates that such models' predictive power may decrease in time, as one may reasonably expect. Nonetheless, we do observe encouraging results over the entire simulated time frame, particularly in certain areas. We first assess counties comprising the Atlanta metropolitan area. The heavily populated counties of Dekalb, Gwinett, Clayton, and Cobb show very good agreement with the simulated data. The most populous county in Georgia, Fulton County, is over-predicted. On the fringe of the Atlanta metropolitan area, Forsyth, Clarke, and Cherokee counties were also over-predicted.
\par Outside of the Atlanta area, we found a general underestimation of the outbreak severity in the Bibb, Richmond, and Chatham counties. As the ``first wave" of infections is well-simulated, this likely represents a shortcoming of the diffusive model. In particular, after the initial re-openings, additional source terms may need to be included in certain areas to model the effects of travelers arriving from elsewhere, as our model does not account for nonlocal movements. In contrast, Dougherty County, which was heavily impacted early in the epidemic, shows the opposite trend, and the outbreak during the later stages of the simulated period is less severe than as predicted by the model. This shows that the model is sensitive to the initial conditions, and spatially-variable parameters, in particular contact rates, may be needed to account for such phenomena properly.


\subsection{Rio de Janeiro, Brazil}\label{RJ_section}

\alv{The last simulation aims at reproducing the COVID-19 outbreak in the state of Rio de Janeiro, Brazil. Rio de Janeiro has the second number of deaths and infected in Brazil, but the largest number of deaths/inhabitant as per January 2021\footnote{\url{https://www.statista.com/statistics/1107109/brazil-coronavirus-deaths-state/}}}

\subsubsection{Model Construction}

We have obtained the map of the state of Rio de Janeiro (RJ) along with the county boundaries in shapefile format from 
\url{ftp://geoftp.ibge.gov.br/organizacao_do_territorio/malhas_territoriais/malhas_municipais/municipio_2017/UFs/}. To triangulate the RJ region, we follow the same steps explained in the previous section.

\begin{figure}[htpb]
    \centering
    \includegraphics[width=0.9\linewidth]{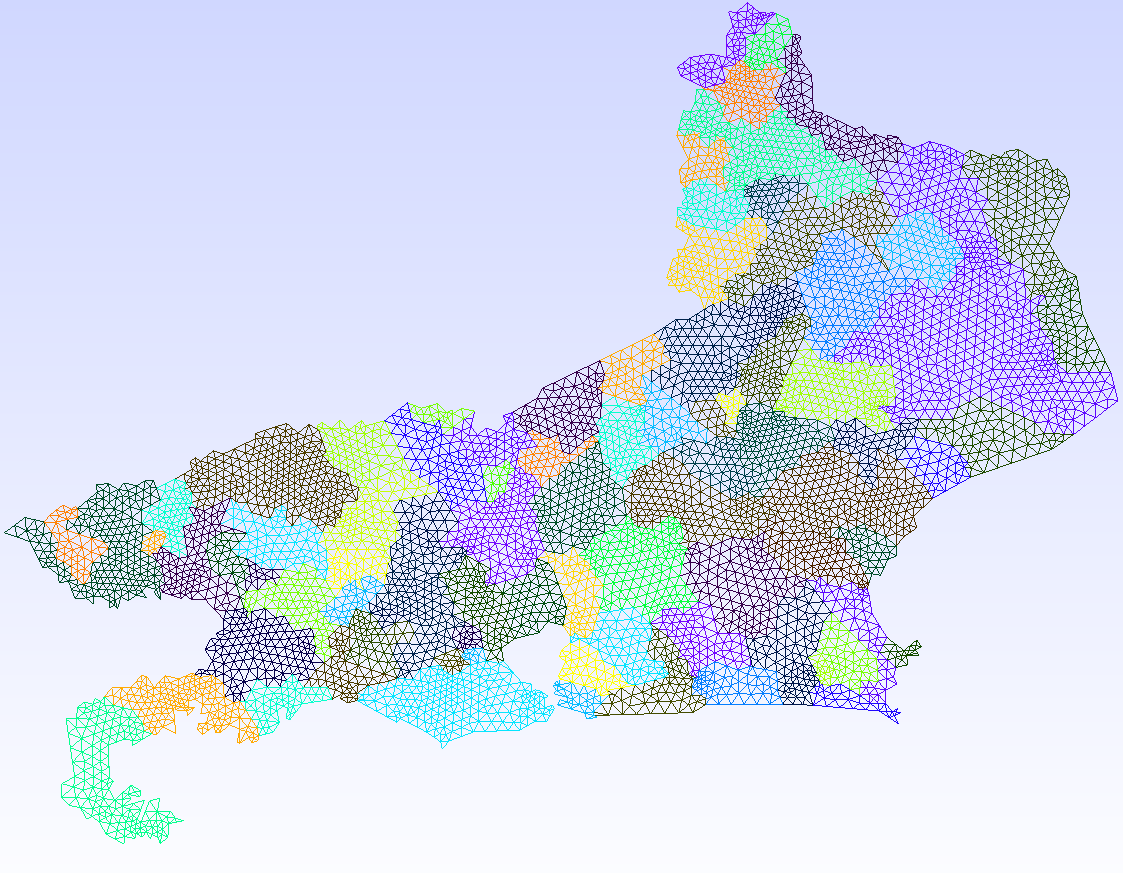}
    \caption{Map of the state of RJ partitioned into 92 internal counties.}
    \label{fig:RJmesh}
\end{figure}

In Fig. \ref{fig:RJmesh}, we show the \alv{grid generated for} the state of Rio de Janeiro. We eliminated some islands to facilitate the simulation. The original mesh has 11,632 linear triangular elements, and after refinement, the minimum spatial resolution is about 500 meters. We initially refine the domain in one level. For the AMR/C procedure, we set $h_{max}=1$, $r_f=0.95$, $c_f=0.01$. We apply the adaptive mesh refinement every 4 time-steps. The time-step is defined as $\Delta t = 0.25$ $days$.

We define the initial infected population accordingly to the data provided by the Brazilian Ministry of Health, collected by \url{https://covid19br.wcota.me/}. We define the beginning of the simulation on 25 March 2020 and simulate 180 days. However, since there is a delay until people develop symptoms, another delay in receiving the test results, and another one for the case being disclosed, we consider that the initial infected population is equal to the case numbers provided for each county on 1 April 2020. As in the case of Georgia, we initialize the exposed population as 10 times the number of infected \cite{reis2020characterization}. The susceptible population is based on the estimation of the population of each county, given by the Brazilian Institute for Geography and Statistics\footnote{\url{https://www.ibge.gov.br/cidades-e-estados/rj.html}}. The populations are divided by the area of each county and distributed on the 92 areas of the mesh as $people/km^2$.

\begin{figure}[htpb]
    \centering
    \includegraphics[width=\linewidth]{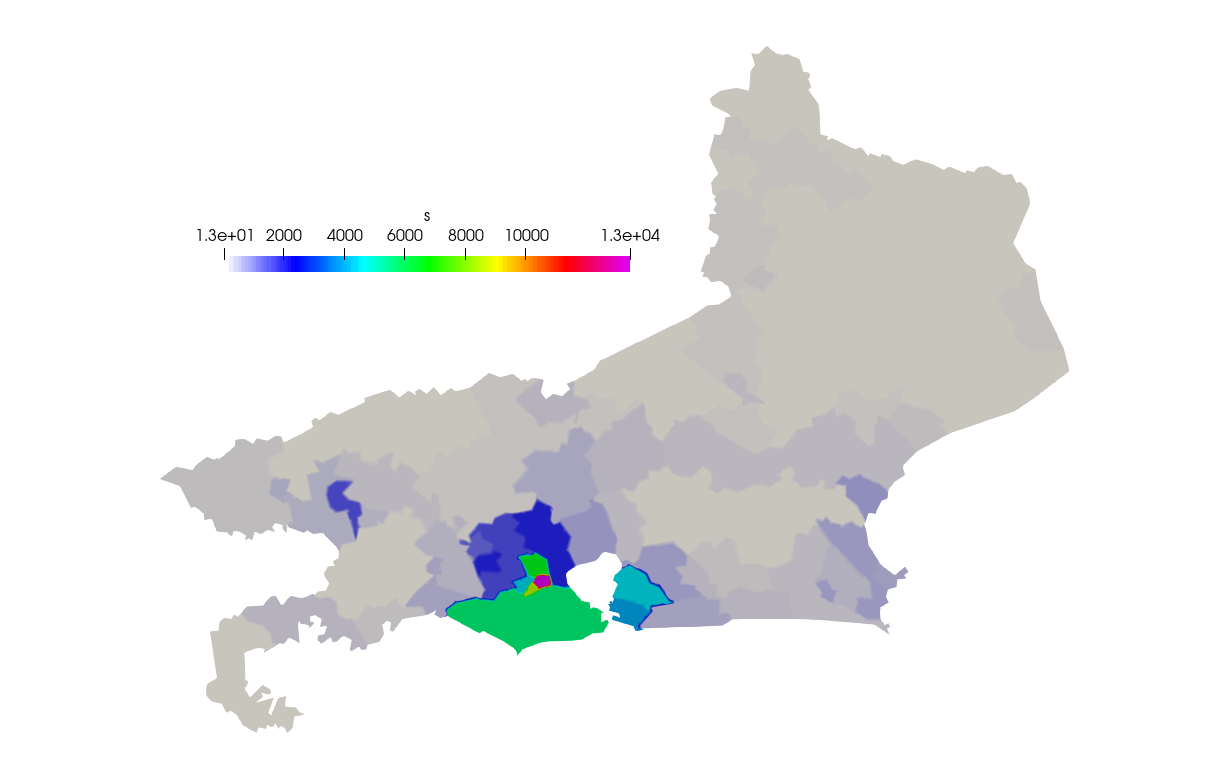}
    \caption{Initial susceptible population ($people/km^2$).}
    \label{fig:RJs}
\end{figure}

\begin{figure}[htpb]
    \centering
    \includegraphics[width=\linewidth]{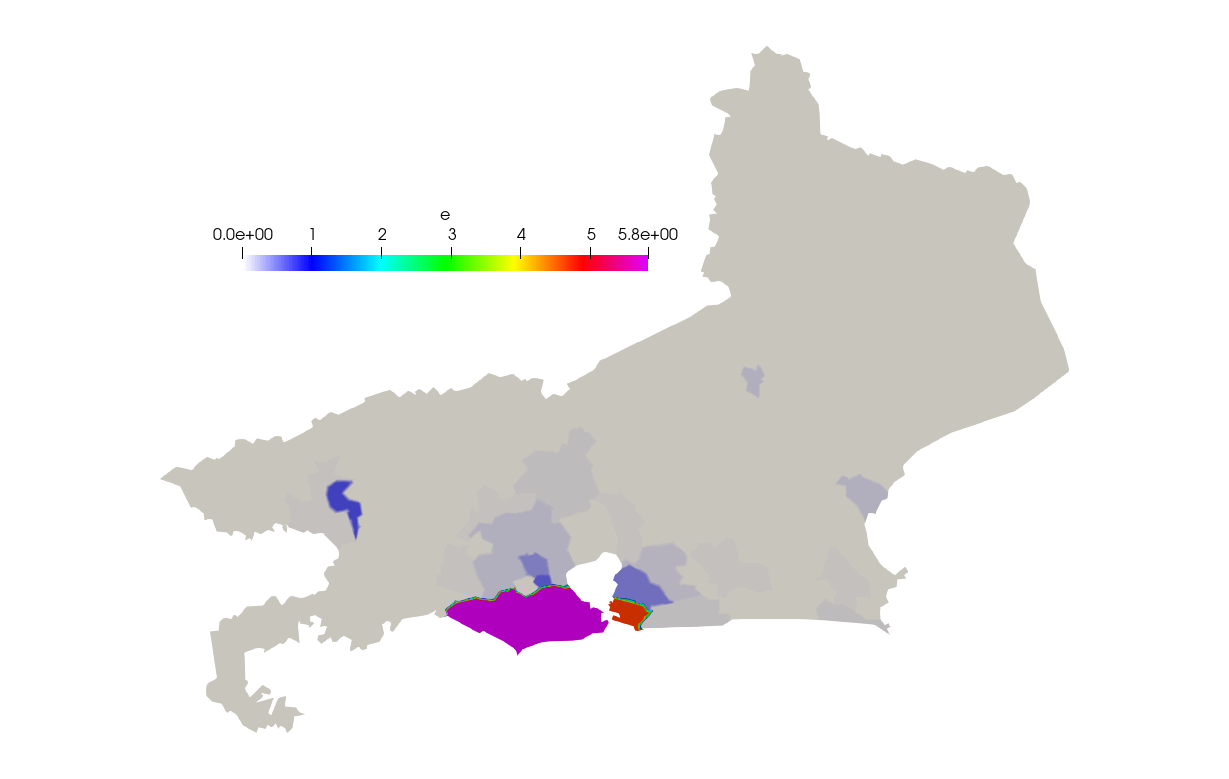}
    \caption{Initial exposed population ($people/km^2$).}
    \label{fig:RJe}
\end{figure}

\begin{figure}[htpb]
    \centering
    \includegraphics[width=\linewidth]{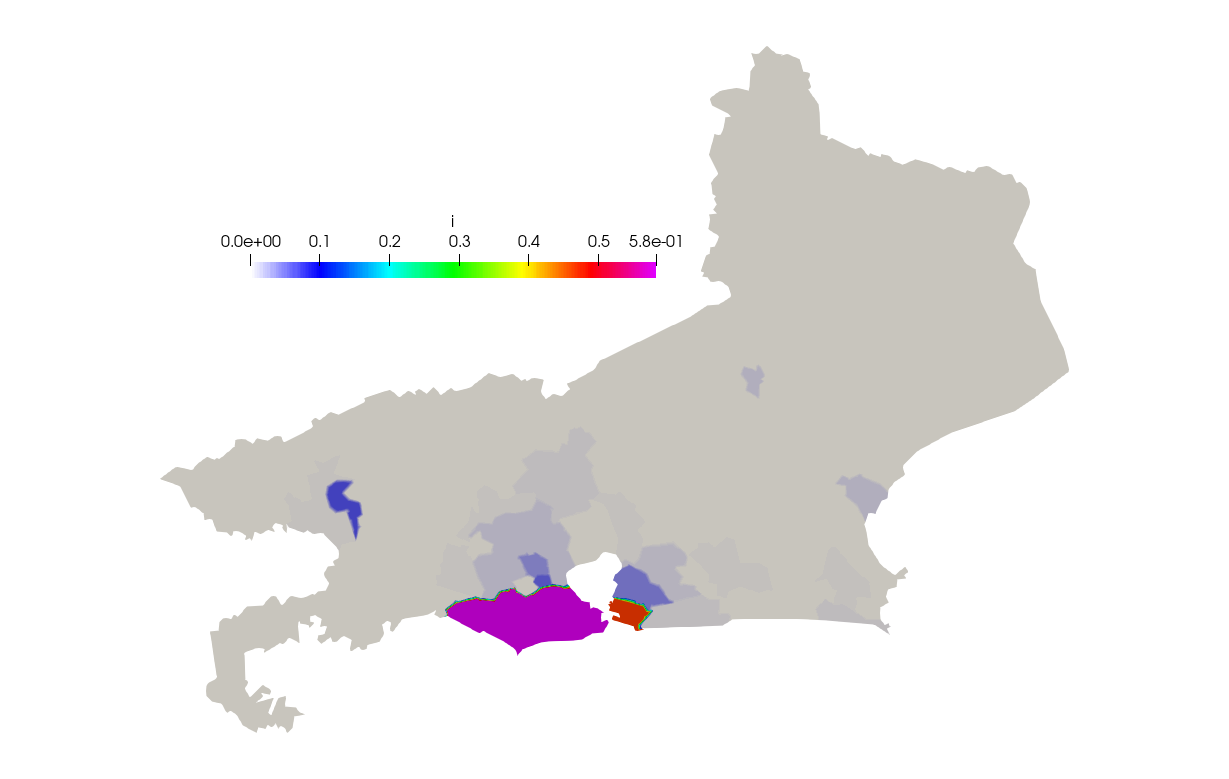}
    \caption{Initial infected population ($people/km^2$).}
    \label{fig:RJi}
\end{figure}

In Figs. \ref{fig:RJs}, \ref{fig:RJe}, and \ref{fig:RJi} we show the initial conditions. The population of the state of Rio de Janeiro is concentrated near the capital and the metropolitan region.


The parameters of the simulation are defined as: $\alpha = 1/7$ $ day^{-1}$, $\gamma_i = 1/24$  $ day^{-1}$,  $\gamma_e = 1/6$  $ day^{-1}$, $\delta = 1/160$  $ day^{-1}$. These values are based on available data from the literature regarding the mortality, incubation period, and recovery time for
infected and asymptomatic patients \cite{viguerie2020simulating}.

\begin{figure}[htpb]
  \centering
  \subfloat[Function for $\nu$.]{\includegraphics[width=0.45\textwidth]{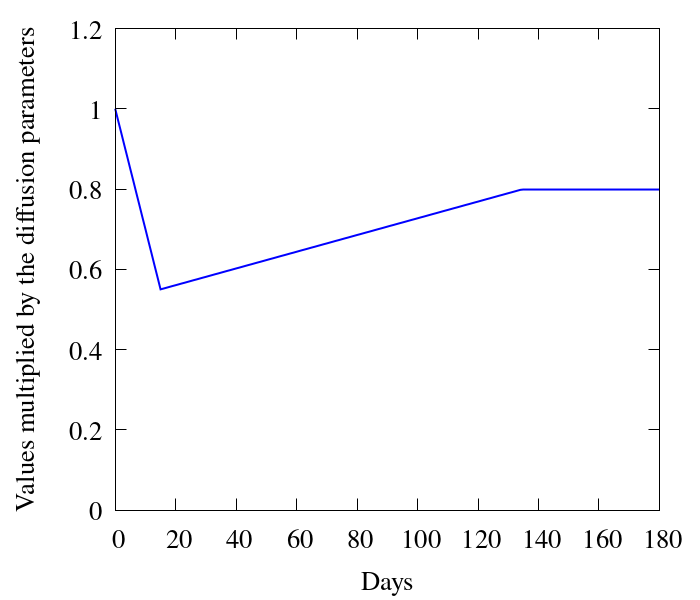}\label{fig:diff_function}}
   \hfill
  \subfloat[Function for $\beta$.]{\includegraphics[width=0.45\textwidth]{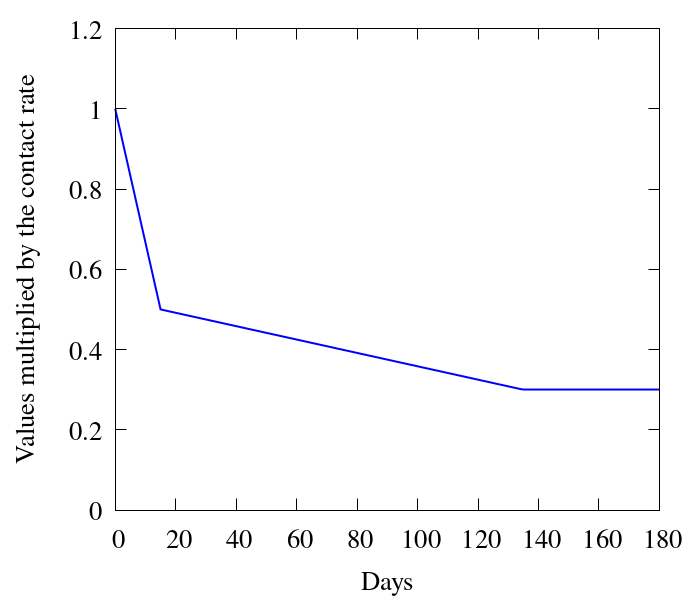}\label{fig:rt_function}}
    \hfill
  \caption{Functions that multiplies the diffusion (left) and contact rate (right) in time for Rio de Janeiro.}
  \label{comp_RJ_1}
\end{figure}

The contact rate is estimated based on the reproduction number estimation given in \url{https://perone.github.io/covid19analysis/}. We define $\beta_e = \beta_i = 0.215$ $days^{-1}$, and we multiply this value by the \alv{time function} given by Fig. \ref{fig:rt_function}, which is a simplification of the behavior of the reproduction number between 25 March 2020 and 21 September 2020. The diffusion coefficient is based on the social distancing estimation given by \url{https://coronavirus.ufrj.br/covidimetro/}, which presents the perception of average weekly social isolation based on the movement of cell phones in the state of RJ. The diffusion behaves opposite to the social isolation, i.e., when one increases, the other one decreases. Therefore, we set $\nu_s = \nu_e = \nu_r = 1\times 10^{-3}$ and $\nu_i = 1\times 10^{-5}$ $km^2persons^{-1}days{ ^-1}$, and scale these values by a \alv{time-dependent function} (Fig. \ref{fig:diff_function}) that tries to represent the social isolation during the 180 days of simulation.

\subsubsection{Results}

\begin{figure}[htpb]
    \centering
    \includegraphics[width=0.6\linewidth]{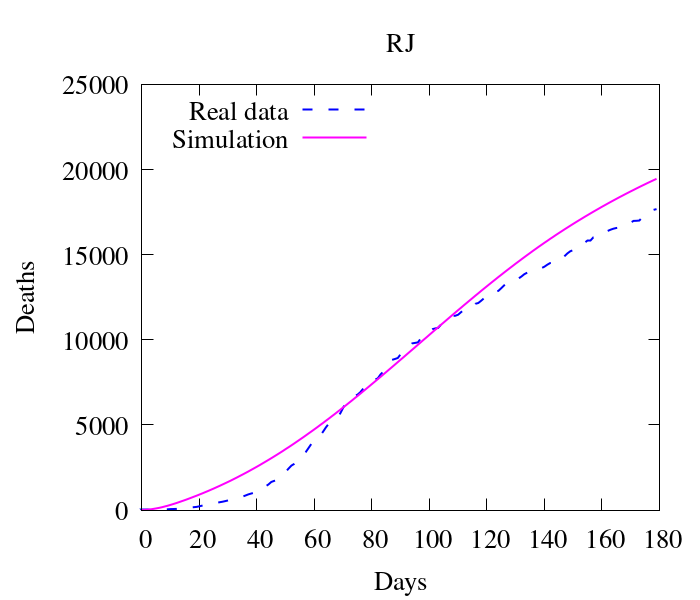}
    \caption{Comparison between simulation and real data of deaths at Rio de Janeiro state (total).}
    \label{d_RJ}
\end{figure}

\begin{figure}[htpb]
  \centering
      \subfloat{\includegraphics[width=0.45\textwidth]{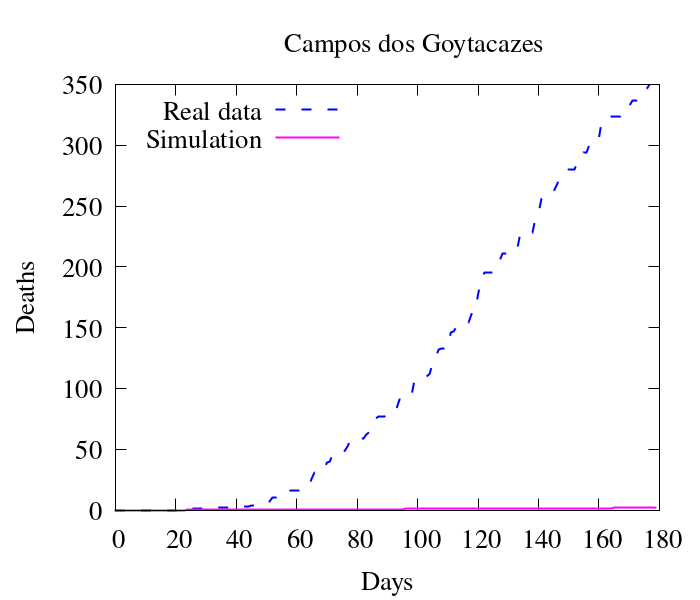}\label{Goytacazes}}
  \hfill
    \subfloat{\includegraphics[width=0.45\textwidth]{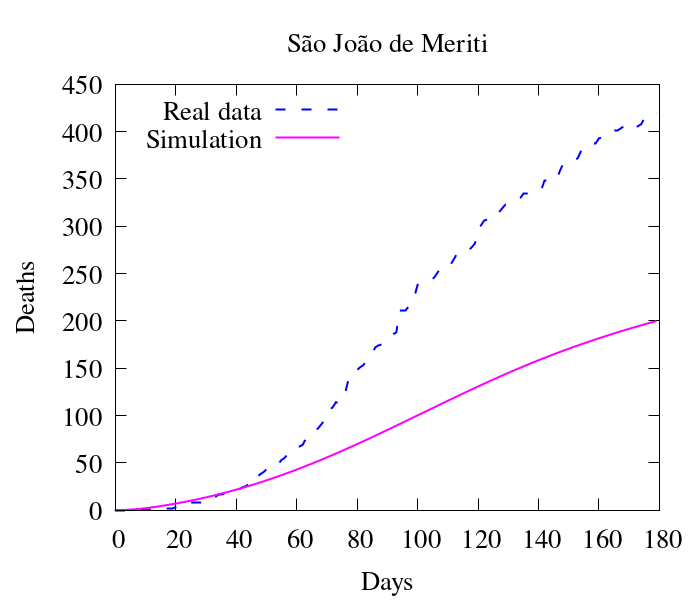}\label{SaoJoaoMeriti}}
        \hfill
  \subfloat{\includegraphics[width=0.45\textwidth]{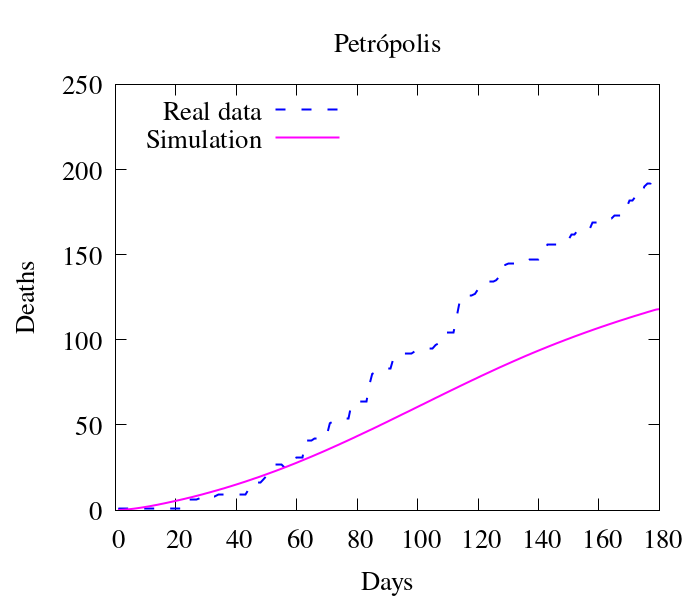}\label{Petropolis}}
  \hfill
  \subfloat{\includegraphics[width=0.45\textwidth]{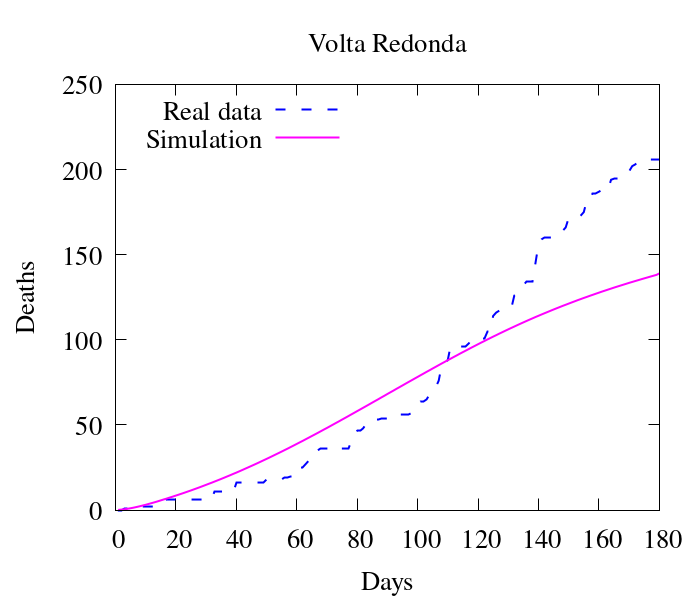}\label{Volta}}
          \hfill
  \subfloat{\includegraphics[width=0.45\textwidth]{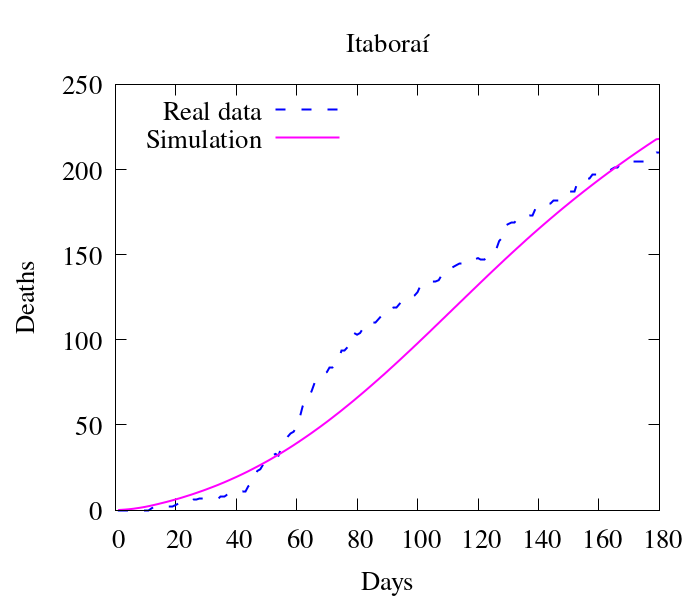}\label{Itaborai}}
  \hfill
  \subfloat{\includegraphics[width=0.45\textwidth]{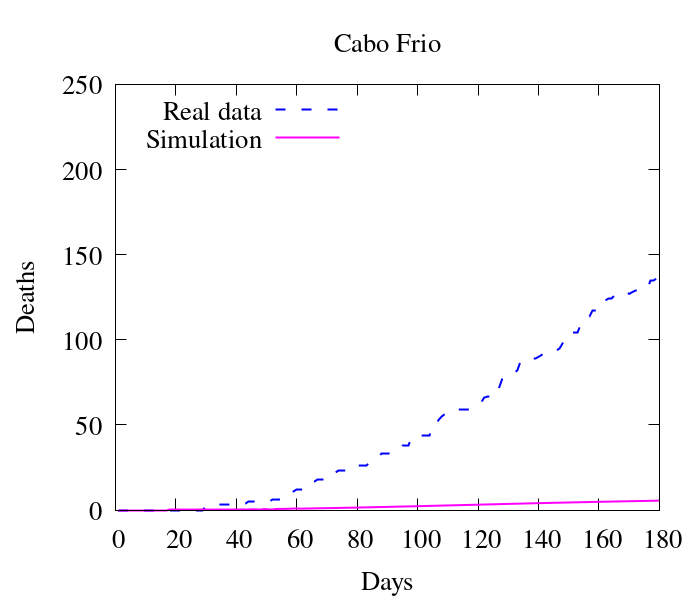}\label{Cabo}}

  \caption{Comparison between simulation and real data of deaths at RJ (per county) - part 2.}
  \label{comp_RJ_2}
\end{figure}

\begin{figure}[htpb]
    \centering
    \includegraphics[width=\linewidth]{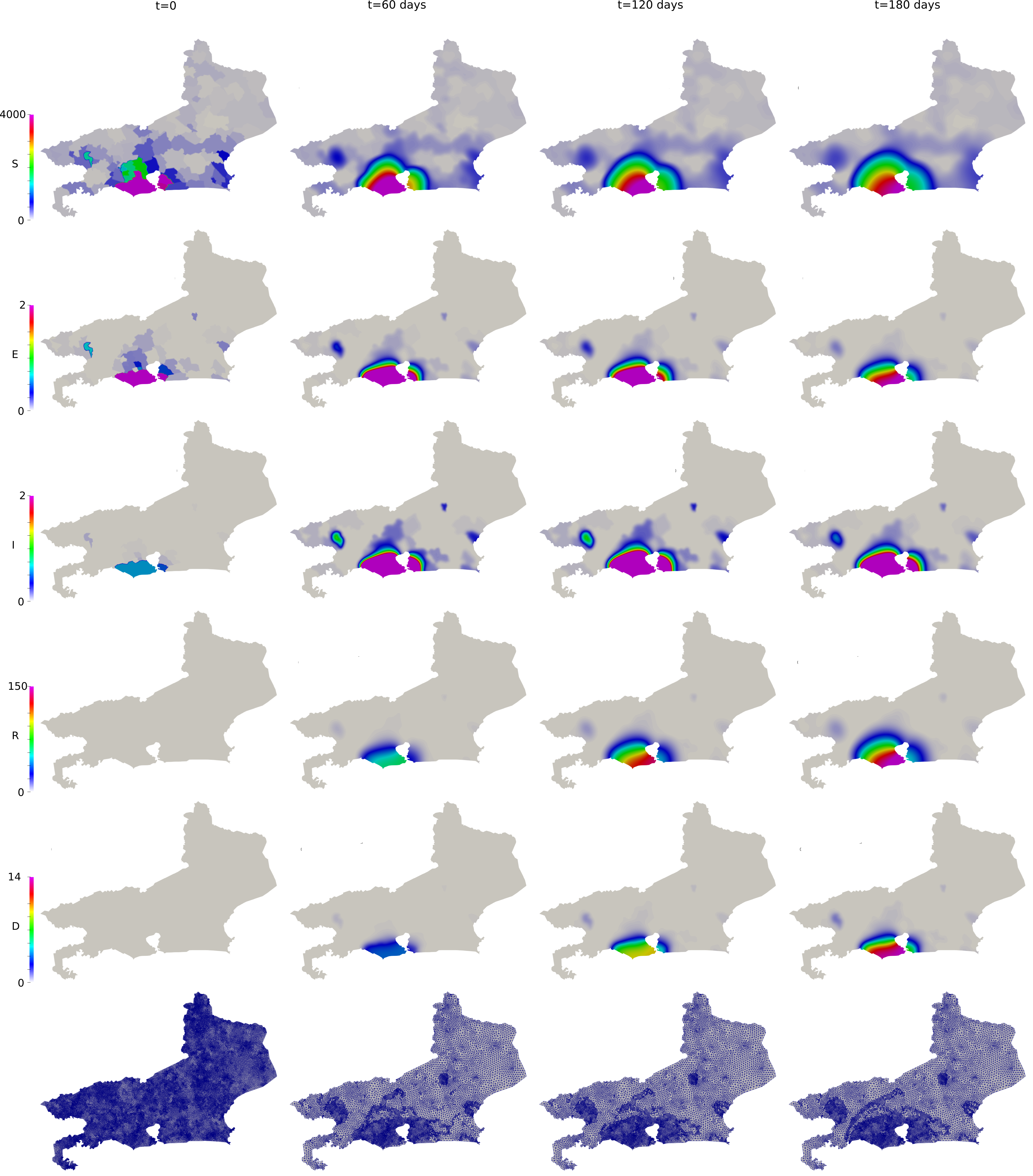}
    \caption{Snapshots of Rio de Janeiro simulation. Top to bottom: Susceptible, Exposed, Infected, Recovery and Deceased population ($people/km^2$), followed by the mesh at different time-steps.}
    \label{fig:rio_simulation}
\end{figure}

To compare the results of our model with the data available, we compare the number of deaths caused by the COVID-19. First, we show the values of the whole state (Fig. \ref{d_RJ}). Then, we integrate the values of some counties and plot them in time (Figs. \ref{comp_RJ_1} and \ref{comp_RJ_2}). Fig. \ref{fig:rio_simulation} also shows some snapshots of the simulation with the populations and mesh at different time-steps.

\begin{figure}[htpb]
  \centering
  \subfloat{\includegraphics[width=0.45\textwidth]{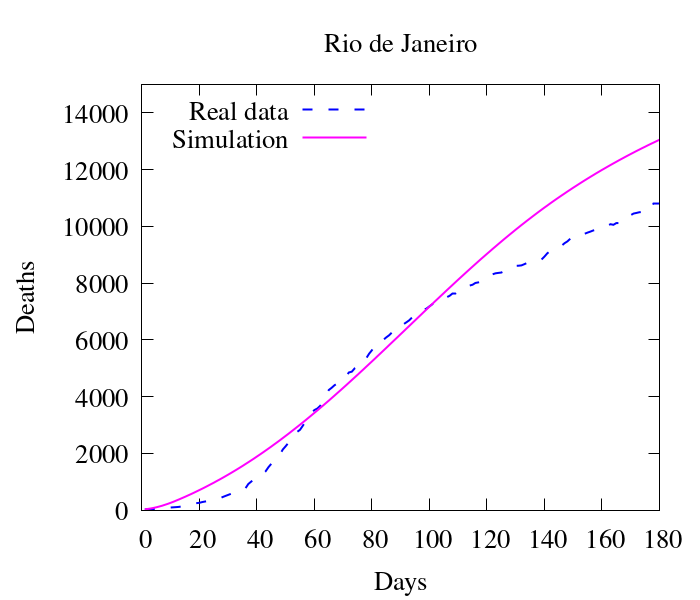}\label{Rio}}
   \hfill
  \subfloat{\includegraphics[width=0.45\textwidth]{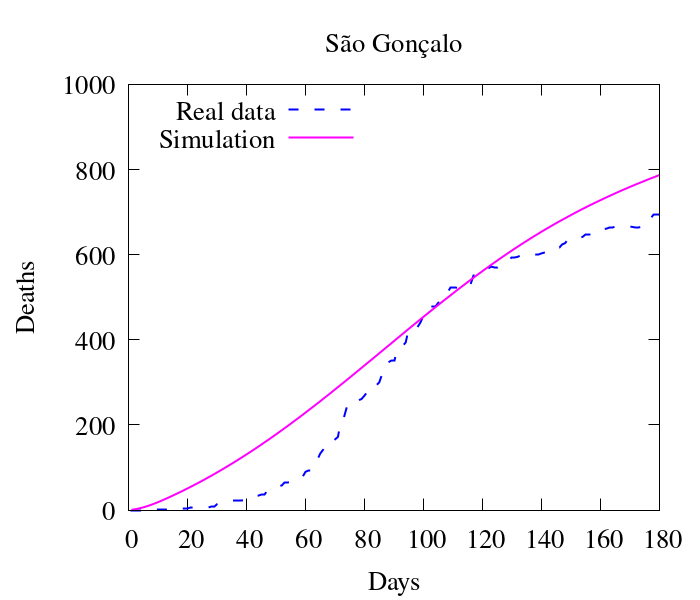}\label{SaoGoncalo}}
    \hfill
  \subfloat{\includegraphics[width=0.45\textwidth]{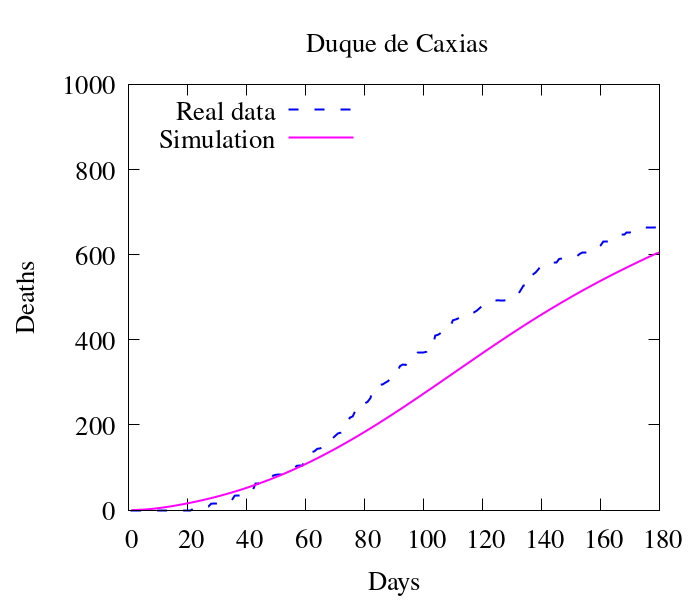}\label{Duque}}
    \hfill
  \subfloat{\includegraphics[width=0.45\textwidth]{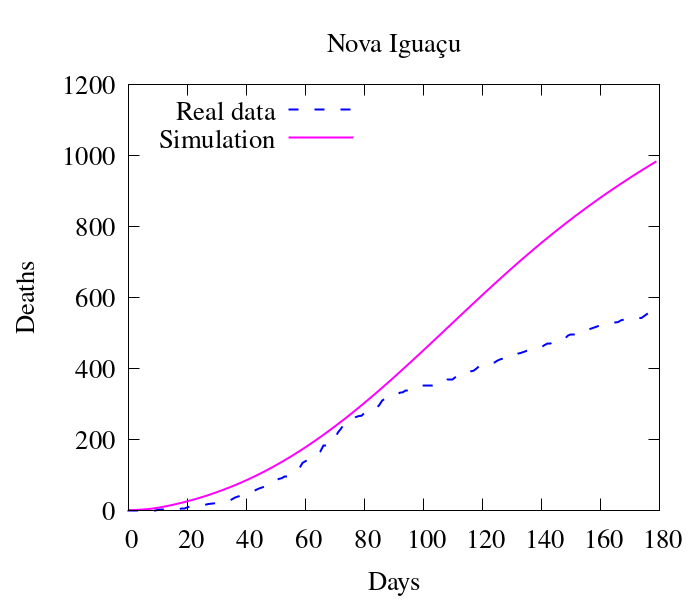}\label{NovaIguacu}}
    \hfill
  \subfloat{\includegraphics[width=0.45\textwidth]{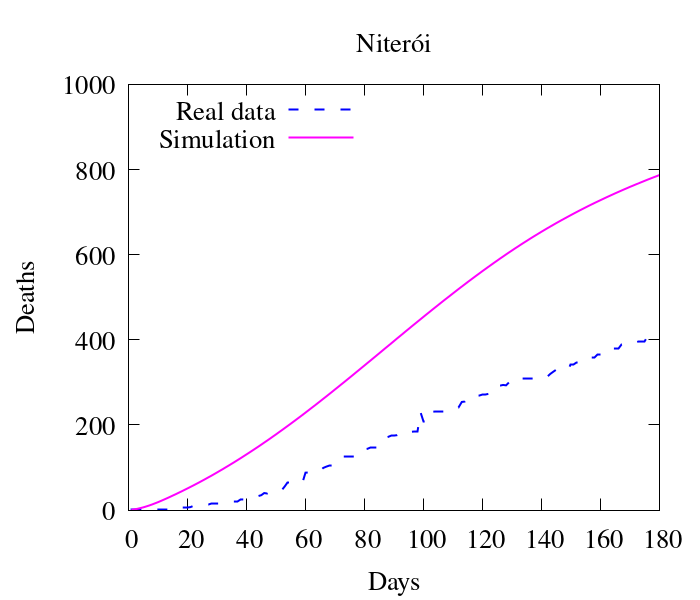}\label{Niteroi}}
  \hfill
    \subfloat{\includegraphics[width=0.45\textwidth]{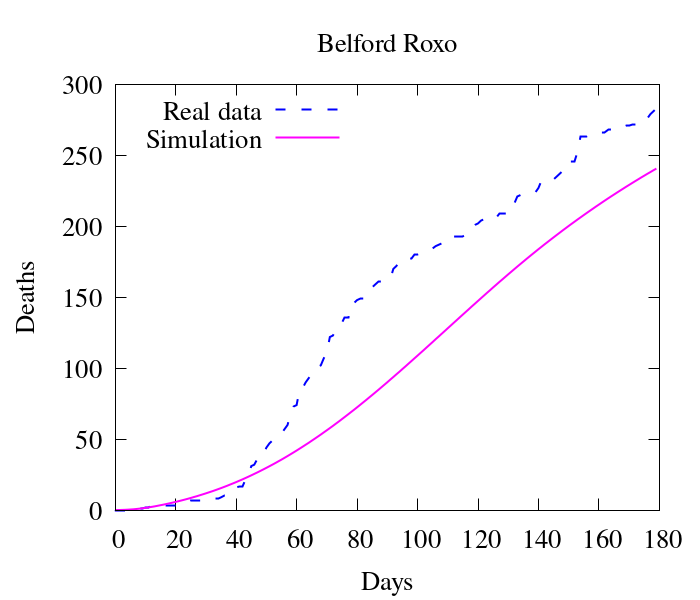}\label{Belford}}
  \caption{Comparison between simulation and real data of deaths at RJ (per county) - part 1.}
  \label{comp_RJ_1}
\end{figure}

As in the previous case in Georgia, most selected counties show good agreement with the available data. \alex{Notably, over the entire region, the relative difference in $L^2$ norm between the observed and simulated fatalities is only 10.06\%. Over the first several weeks, the results are in good agreement nearly everywhere, and, in the areas around the city of Rio de Janeiro, we capture the dynamics very accurately over the entire considered time interval. In addition to the county of Rio de Janeiro, we observe particularly good agreement in the highly populated areas of São Gonçalo, Duque de Caxias, Belford Roxo, and Itaborai. In the areas of Petropolis, Volta Redonda, and Nova Igauçu, we obtain somewhat less accurate, but still nonetheless reasonable results. In these latter cases, we observe good agreement over the initial phases of the dynamics, with some differences later in the time interval. These deviations are unsurprising, as one may naturally expect increased difficulty further in time. } 
\par \alex{We do observe some notable instances of overestimation in Niterói and underestimation in Cabo Frio and Campos dos Goytacazes.} \alex{In the latter cases, there were no initial infected or exposed population in this simulation, making diffusion the only possible way for the virus to reach the areas. Given the large distances between these areas and the hotspots near the city center of Rio de Janeiro (157 km to Cabo Frio and 279 km to Campos dos Goytacazes), as well as large sparsely-populated areas between them, population-weighted diffusion was not able to properly represent these dynamics. The model best captures spatial dynamics over relatively short distances overpopulated regions. This represents a shortcoming of the model, and additional terms, such as source/sink terms, fractional diffusion operators, or bilaplacian diffusion terms, may be required to properly account for such nonlocal dynamics.} \alex{In the case of Niterói, the overprediction may result from an overestimation of the initial exposed population or failure to take into account specific local policies. This is similar to what we observed in Georgia, in which certain particular regions had dynamics that deviated from the measured data. However, as in Georgia's case, we observe very good agreement when considering the totality of the region, with particular areas less well-represented due to the natural limitations of our modeling approach. In these instances, such cases provide clear directions in which we may improve the modeling framework.}

\section{Conclusions and future work}

\alex{In this work, we have established the robustness of the SEIRD model introduced in \cite{viguerie2020simulating} and further extended in \cite{viguerie2020diffusion, grave2020adaptive, jha2020bayesian} by examining both its frequency- and density-dependent formulations and applying it to three settings over different continents: the region of Lombardy, Italy (density-dependent), \alv{the U.S. state of Georgia, and the state of Rio de Janeiro, Brazil} (frequency-dependent). All these regions have very different geographic characteristics, population density patterns, policy response measures, and cultural contexts generally; despite this, the models were able to adequately reproduce the spatio-temporal contagion dynamics with minimal parameter differences and very little tuning. This provides strong evidence towards the robustness of the model, as it can produce adequate results across different settings without requiring extensive parameter fitting and learning. }

\alex{The results obtained showed good qualitative agreement generally across all regions; however, they show clear room for improvement. In the cases of \alv{Georgia and Rio de Janeiro,} we find that the initial conditions have a large influence, even over the long-term dynamics of the system. This is particularly apparent in the case of Dougherty County in Georgia, where an early outbreak led to a long-term overestimation of contagion, and in Campos dos Goytacazes and Cabo Frio in Rio de Janeiro, where a relative lack of early cases led to an under-prediction of the long term contagion effects. This is likely due to the diffusion formulation limitations, which cannot correctly account for non-local mobility across different areas, such as returning travelers. Such effects could be incorporated in a number of ways by including more sophisticated source/sink terms, fractional diffusion operators, or bilaplacian terms. In Lombardy, we observe an under-prediction of contagion in less-populated regions; this appears to be characteristic of the density-dependent formulation, and the frequency-dependent models applied in Georgia and Rio de Janeiro do not suffer from this problem. Across all cases, the same parameter values were used everywhere. While this represents evidence towards robustness and is in some sense a positive aspect, results could, of course, be further improved by using different definitions in different regions, something explored briefly in \cite{viguerie2020simulating} and in more detail in \cite{jha2020bayesian}. Such approaches may incorporate more sophisticated methods, including optimization and machine-learning techniques. These shortcomings represent natural directions for future work in this area. Nonetheless, we believe that our results clearly establish the viability and robustness of this PDE modeling framework, suggesting this as a suitable path for future research, and ultimately may perhaps prove useful to public health decision-makers.}



\section{Acknowledgements}
This research was financed in part by the Coordena\c{c}\~ao de Aperfei\c{c}oamento de Pessoal de N\'ivel Superior - Brasil (CAPES) - Finance Code 001 and CAPES TecnoDigital Project 223038.014313/2020-19.  This research has also received funding from CNPq and FAPERJ. A. Reali was partially supported  by the Italian Ministry of University and Research (MIUR) through the PRIN project XFAST-SIMS (No. 20173C478N).

%
%


\bibliographystyle{unsrt}      
\bibliography{references}   

%
%

\end{document}